\documentclass[%
 reprint,
 amsmath,amssymb,
aps
pra,twocolumn,
]{revtex4-1}
\usepackage{braket}
\usepackage{graphicx}
\usepackage{bm}
\usepackage{float}
\usepackage{color}
\usepackage{soul}
\usepackage[caption = false]{subfig}
\usepackage{hyperref}
\usepackage{placeins}

\usepackage[normalem]{ulem}

\begin{document}




\title{Optical squeezing for an optomechanical system without quantising the mechanical motion}
\author{Yue Ma$^1$}

\author{Federico Armata$^1$}

\author{Kiran E. Khosla$^1$}

\author{M. S. Kim$^{1,2}$}

\affiliation{$^1$QOLS, Blackett Laboratory, Imperial College London, London SW7 2AZ, United Kingdom \\
$^2$Korea Institute of Advanced Study, Seoul 02455, South Korea
}

\begin{abstract}

Witnessing quantumness in mesoscopic objects is an important milestone for both quantum technologies and foundational reasons. Cavity optomechanics offers the ideal system to achieve this by combing high precision optical measurements with mechanical oscillators. However, mechanical quantumness can only be established if the behaviour is incompatible with any classical description of an oscillator. After explicitly considering classical and hybrid quantum-classical descriptions of an optomechanical system, we rule out squeezing of the optical field as such a witness by showing it is also predicted without quantizing the mechanical oscillator.

\end{abstract}


\maketitle

\section{introduction}

Witnessing the quantum nature of a physical system is a central and recurrent goal in physics. Quantumness can only be unambiguously demonstrated when predictions based on all possible classical theories are violated~\cite{kot2012breakdown}, for example by the violation of a Bell inequality~\cite{bell1964einstein,vivoli2016proposal,marinkovic2018optomechanical}, or detecting Wigner negativity via tomographic reconstruction~\cite{glauber1963coherent,aspelmeyer2014cavity,jacobs2009engineering,vanner2011selective}. However, for experiments without such unambiguous witnesses, a classical description may predict the observed result. In this case, even a nonclassical state may not necessarily have its nonclassicality revealed.
This has already been pointed out in a variety of systems, including Josephson oscillator~\cite{gronbech2010tomography}, Rabi oscillation in light-atom interaction~\cite{ruggenthaler2009rabi}, optical Berry phase~\cite{haldane1987comment,segert1987photon}. Here, and in the following we use the word ``quantumness'' to mean the \emph{necessity} of quantisation given certain experimental setups, irrespective of any nonclassical features in the state as described by quantum mechanics.

Several general, operational criteria for characterizing quantumness have been proposed~\cite{mari2011directly,kot2012breakdown,alicki2008simple,richter2002nonclassicality}, most of which are directly applicable to photonics where one has direct access to the relevant mode. However, one would also like to study the quantumness criterion of a mode without direct access, optomechanics being a typical example~\cite{aspelmeyer2014cavity}. Optomechanics uses high precision control of optical or microwave fields to manipulate and readout the micromechanical motion of massive oscillators, and has been widely studied for both foundational~\cite{pikovski2012probing,penrose1998quantum,marshall2003towards,penrose1996gravity,diosi1989models,schmole2016micromechanical,plato2016gravitational,latmiral2016probing} and practical applications~\cite{gavartin2012hybrid,forstner2012cavity,krause2012high,bagci2014optical,vanner2011pulsed,vanner2015towards}. Characterising which aspects of the optical field can demonstrate the quantumness of the mechanical oscillator is of foundational interest. This has been briefly investigated in optomechanics in the context of tracing the origin of experimentally observed sideband asymmetry~\cite{safavi2012observation,weinstein2014observation,sudhir2017appearance}. A direct theoretical comparison between quantum and classical descriptions of the optomechanically generated phase has been considered in light of interferometric experiments~\cite{armata2016quantum}. 

Here, we focus on optomechanically generated squeezing of a single mode optical field~\cite{fabre1994quantum,mancini1994quantum}, previously observed in several experiments~\cite{brooks2012non,purdy2013strong,safavi2013squeezed}, and investigate whether optical squeezing is a signature for mechanical quantumness. The experiments operate in the linearised regime, and are therefore expected to reveal less quantumness than the nonlinear optomechanical interaction~\cite{dellantonio2018quantum,lloyd1999quantum}. In this work we consider the latter, thereby maximising the possibility for detecting mechanical quantumness in optomechanics. We examine classical, quantum and hybrid quantum-classical theories, focusing on the temporal evolution of the field quadrature variance. We show optical squeezing is predicted without quantising the mechanical oscillator and is therefore eliminated as a mechanical quantumness witness, regardless of the oscillator temperature. We emphasise the aim of the work as ruling out optical quadrature squeezing as evidence for mechanical quantumness, instead of proposing a new experimental witness for it.

\section{Quantum description}

We examine a closed optomechanical system with the intracavity field initialised to a coherent state. The joint system undergoes unitary evolution and we compute the time evolution of the intracavity field variance. Note this setup is different from most experiments~\cite{aspelmeyer2014cavity}, where the cavity is driven by an external laser, the output field from the cavity is monitored, and the system is subject to dissipation. Laser driving and dissipation introduce open system dynamics, smearing out any mechanical quantumness signature. We neglect these elements to maximise the mechanical quantumness effect in optical field variance.

\begin{figure*}[t]
\centering
\includegraphics[width=\textwidth]{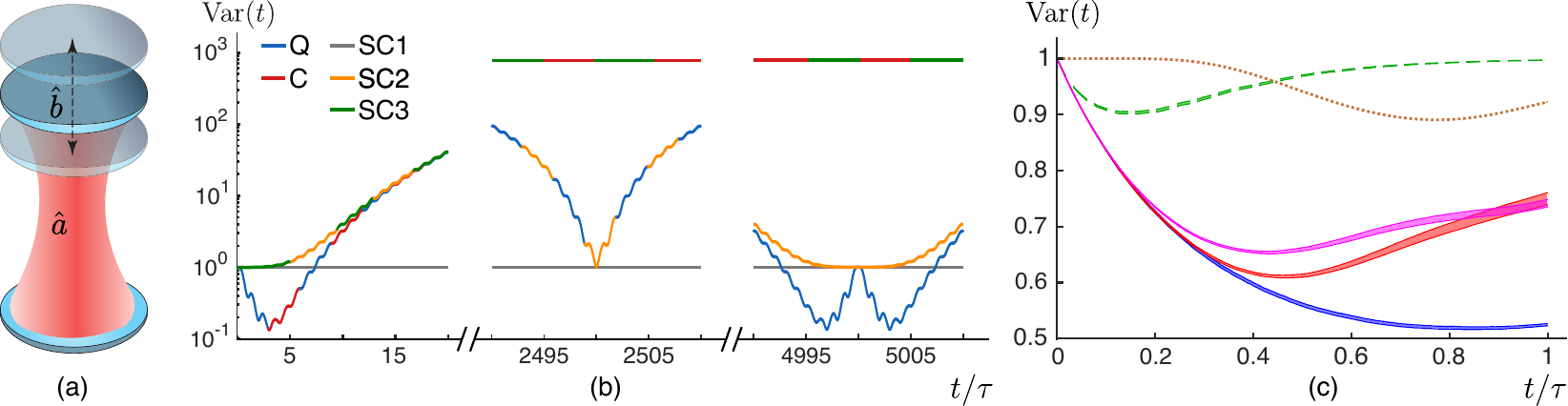}
\caption{(a) A Fabry-P\'{e}rot cavity with one movable mirror described as a harmonic oscillator. (b) Field variance as a function of time for quantum (blue, ``Q") and classical (red, ``C") description, mean field hybrid descriptions with constant $I$ (grey, ``SC1"), Poisson random $I$ (orange, ``SC2"), and Gaussian random $I$ (green, ``SC3"). All plots use $\alpha=20$, $k=0.01$. A line containing two alternating colours means the two corresponding descriptions predict indistinguishable results. (c) Field variance as a function of time for the hybrid measurement model with $k=0$ (blue line), $k=0.1$, $x(0)=0$, $p(0)=0$ (red line), $k=0.1$, $x(0)$ and $p(0)$ as classical random variables simulating the zero-point-fluctuation (pink line). All curves use $\alpha=2$, $\Gamma=0.01\omega$. For comparison, the effect of cavity dissipation is shown for the hybrid model (green dashed line) with photon dissipation rate $\kappa=\omega$, and the quantum model (brown dotted line) with $\kappa=0.3\omega$. Linewidth for numerical solutions represents $\mathrm{mean}\pm\mathrm{standard~deviation}$.
}\label{fig:1}
\end{figure*}

We consider a Fabry-P\'{e}rot cavity with a movable mirror, shown in Fig.~\ref{fig:1}(a), the equilibrium frequency of the oscillator (field) given by $\omega$ $(\Omega)$. The Hamiltonian in the frame rotating with the optical frequency $\Omega$ is~\cite{pace1993quantum,law1995interaction}
\begin{equation}
\hat{H}/\hbar=\omega\hat{b}^{\dagger}\hat{b}-\frac{g_0}{\sqrt{2}}\hat{a}^{\dagger}\hat{a}(\hat{b}^{\dagger}+\hat{b}),
\label{eq:H_q}
\end{equation}

\noindent where $\hat{a}~(\hat{b})$ is the optical (mechanical) annihilation operator and $g_0/\sqrt{2}$ is the single-photon optomechanical coupling strength. The time evolution of the field in the Heisenberg picture is~\cite{mancini1997ponderomotive}
\begin{equation}
\hat{a}(t)=e^{iA(t)/2}e^{k\big((1-e^{-i\omega t})\hat{b}-(1-e^{i\omega t})\hat{b}^{\dagger}\big)}e^{iA(t)\hat{a}^{\dagger}\hat{a}}\hat{a},
\label{eq:quantum}
\end{equation}
where $A(t)=2k^2(\omega t-\sin\omega t)$ and $k=g_0/(\sqrt{2}\omega)$. In the following we will be interested in the $\theta$-dependent quadrature operator $\hat{\sf X}_{\theta}(t)=\hat{a}(t)e^{-i\theta}+\hat{a}^{\dagger}(t)e^{i\theta}$, and the corresponding variance ${\rm Var}_{\theta}(t)=\langle\hat{\sf X}_{\theta}^2(t)\rangle-\langle\hat{\sf X}_{\theta}(t)\rangle^2$. Squeezing is defined based on the minimum of variance taken over all quadrature angles, $\mathrm{Var}(t)\equiv\min_{\theta}\{\mathrm{Var}_{\theta}(t)\}$. If $\mathrm{Var}(t)<1$, the field is squeezed at time $t$. Note the effect of the mechanical commutation relation only comes in as the phase $e^{iA(t)/2}$, thus not able to change $\mbox{Var}(t)$.

The intracavity field is initialised to a coherent state $|\alpha\rangle_{\rm L}$ with $\alpha$ real (without loss of generality), and the mechanical oscillator to a vacuum state (for generalization to initial thermal states see Appendix). The time evolution of ${\rm Var}(t)$ is shown in blue in Fig.~\ref{fig:1}(b) (see the analytical expressions in Appendix). For simplicity we only plot the nontrivial parts where different descriptions to be considered later predict distinguishable variances. We choose $\alpha=20$, with strong optomechanical interaction $k=0.01$, similar to parameters predicted in the high cooperativity regime of thin-film superfluid~\cite{baker2016theoretical} (for other parameters see Appendix). The field is squeezed for the first few mechanical periods. Then the variance rapidly increases to a stable value $2\alpha^2+1$ as expected. At approximately $t=\tau/4k^2$, where $\tau=2\pi/\omega$ is the mechanical period, there is a quantum revival (half revival in the form of optical cat state when $t$ is integer multiple of $\tau$) where variance rapidly returns to $\gtrsim1$ for several mechanical periods before again stabilising at $2\alpha^2+1$. There is another qualitatively different revival after another interval of $\Delta t\approx\tau/4k^2$ (where $|\alpha\rangle\rightarrow|-\alpha\rangle$). The quantum variance again rapidly decreases, this time below $1$, indicating optical squeezing. The two quantum revivals are repeated periodically. The behaviours are similar to the time evolution of phase space distribution of an optical field in a Kerr medium~\cite{milburn1986quantum}, as Hamiltonian Eq.~\eqref{eq:H_q} can be understood as (an oscillator-mediated) intensity-dependent optical phase shift~\cite{james2007effective}. Note at the end of each mechanical period, the field and oscillator decouple~\cite{armata2016quantum}. The field variance is independent of the oscillator state, thus the appearance of squeezing and revival are independent of the mechanical temperature (see Appendix).

The quantum description will serve as the reference with which alternative classical or hybrid descriptions will be compared. Irreproducible field variance behaviours by other descriptions are candidates of mechanical quantumness witness.

\section{Classical description}

The initial states for both the field and the oscillator have well-defined positive phase space distributions. Classical ensemble dynamics can thus be defined by averaging over different time evolutions where the initial conditions are sampled from probability distributions matching the initial quantum states.

The Hamiltonian Eq.~\eqref{eq:H_q} is mapped into a classical Hamiltonian, with classical canonical field (oscillator) variables $\alpha_{\mathrm{L}},\alpha_{\mathrm{L}}^{*}~(x,p)$. Classical uncertainty in the field (oscillator) is described by a probability distribution of $\alpha_{\mathrm{L}}$ ($x$ and $p$) over phase space, representing the ignorance of the exact state of the system~\cite{marshall1988stochastic}. Specifically, we choose the initial field amplitude $\alpha_{\rm L}(0)=\alpha+\delta$, where $\alpha$ is the coherent state amplitude $\ket{\alpha}_{\rm L}$, $\delta$ is a complex zero mean Gaussian random variable with covariance matrix $\rm diag(1/2,1/2)$ to classically simulate the vacuum noise. The initial oscillator position and momentum satisfy the Maxwell-Boltzmann distribution such that the initial classical variance matches the quantum ground state variance.

Time evolution of the classical field amplitude is solved from Hamilton's equations (see Appendix),
\begin{equation}
\alpha_{\mathrm{L}}(t)=e^{i\sqrt{2}k\big(x(0)\sin\omega t+p(0)(1-\cos\omega t)\big)}e^{iA(t)|\alpha_{\mathrm{L}}(0)|^2}\alpha_{\mathrm{L}}(0),
\label{eq:classical}
\end{equation}
depicting a deterministic time evolution of a closed system for a given realization of $\delta$, $x(0)$ and $p(0)$. The time evolution of field variance is derived by integrating out the random initial conditions using the appropriate probability densities. The result is shown as the red line in Fig.~\ref{fig:1}(b). The classical variance is nearly identical to the quantum variance during the initial evolution, as long as $\alpha^2\gg1$ and $k^2\ll1$, applicable to state of the art optomechanical experiments. The noise reduction below its initial reference value acts as the classical counterpart of optical squeezing defined in terms of quadrature variance. As a result, merely observing reduction of optical quadrature noise is not sufficient for claiming mechanical quantumness.



The difference between quantum and classical descriptions appears at the first quantum revival. The classical variance does not show a revival. Each point in the continuous phase space rotates at a different, radial-dependent rate, smearing the initial phase space distribution around a circle of radius $\approx \alpha$. 
The detection of quantum revivals can distinguish quantum and classical descriptions so far. The minimum condition to observe the first revival of squeezing, which requires the field has not leaked out from the cavity by the time of the revivals, is $g_0^2/(\omega\kappa)>1$, where $\kappa$ is the cavity field dissipation rate. This stringent constraint is known as the single-photon blockade condition~\cite{aspelmeyer2014cavity}, and is significantly stronger than strong single photon cooperativity, $\mathcal{C}_0=2g_0^2/(\kappa\gamma)>1$, with $\gamma$ the mechanical dissipation rate. Single photon blockade regime has not yet been achieved in optomechanical experiments with the possible exception of ultracold atoms~\cite{murch2008observation,brennecke2008cavity}. Note mechanical damping and optical decay will further reduce the visibility of quantum revivals (see Appendix).

A fully classical description is not necessarily the optimal choice as we are more interested in quantumness of the oscillator itself instead of the joint optomechanical system. In the following sections we will consider hybrid models consisted of a quantum field and a classical oscillator.

\section{Mean field approximation}

We further clarify signatures of mechanical quantumness by only quantising the field, leaving the oscillator classical. Fundamentally there is an incompatibility when dynamically coupling quantum and classical degrees of freedom, and a unique, unambiguous hybrid description does not exist. Extra assumptions are always required for a self-consistent theory~\cite{eppley1977necessity,diosi2000quantum,adler1994generalized,elze2012linear}. In this section we consider the mean field approximation consisting of a quantum interaction Hamiltonian $\hat{H}_{\mathrm{L}}=-g_0x(t)\hat{a}^{\dagger}\hat{a}$ for the quantized field, and a classical Hamiltonian $H_{\mathrm M}=\frac{1}{2}\omega\big(x^2(t)+p^2(t)\big)-g_0Ix(t)$ for the classical oscillator. The dimensionless intensity $I$, acting as the classical counterpart of photon number, is time-independent as $[\hat{H}_{\rm L},\hat{a}^{\dagger}\hat{a}]=0$.

$I$ can be interpreted in different ways depending on the level of the field's ``quantumness" seen by the classical oscillator. Here we consider three intuitive possibilities although we note there are many more. Firstly, the oscillator is perturbed by only the mean field, hence $I$ is given by the standard mean field $I=\langle\alpha|\hat{a}^{\dagger}\hat{a}|\alpha\rangle$ approximation. Secondly, the oscillator is perturbed by the energy quanta of the field with $I$ given by a random number 
following Poisson distribution with mean and variance $\alpha^2$, coinciding with the photon number distribution for a coherent state. Thirdly, the oscillator can detect fluctuations in the field intensity, but it cannot tell its energy discreteness. $I$ is thus well approximated by a Gaussian distribution with its mean and variance $\alpha^2$, where the approximation holds for $\alpha\gg1$ when the Possion distribtuion of $I$ is well approximated by a Gaussian.

For each case Var($t$) is plotted in Fig.~\ref{fig:1}(b), where the grey, orange and green lines correspond to constant, Poisson and Gaussian $I$, respectively. Constant $I$ results in a small periodic variation above the initial value $1$ at the period of the mechanical oscillation due to the stochastic initial condition of the oscillator. A Poisson distributed $I$ does not exhibit squeezing, but successfully reproduces the periodic revivals at the same times as the quantum description. This highlights that quantum revivals are signatures of the quantisation of the field itself, which guarantees that $I$ only takes integer values. It is not related to whether the oscillator is quantised by representing $x$ and $p$ with $\hat{b}$ and $\hat{b}^{\dagger}$. The existence of revivals further suggests, only if squeezing is observed during quantum revivals can we rule out the hybrid description here, thus claiming the mechanical quantumness. It indicates an even stronger requirement than the single-photon blockade condition. A Gaussian distributed $I$, although acting as a continuous approximation of Poisson distribution, does not show revivals, as the field intensity discreteness is smeared out.

The quantum interaction Hamiltonian $\hat{H}_{\mathrm{L}}=-g_0x(t)\hat{a}^{\dagger}\hat{a}$ only induces a frequency modulation of the single mode optical field without generating squeezing. This is true for any $x(t)$ regardless of the dependence on the field state $\rho(t)$. Consider the evolution of the density matrix in a small time step $\delta t$, $\rho(t+\delta t)=\exp[ig_0\delta t x(t)\hat{a}^{\dagger}\hat{a}]\rho(t)\exp[-ig_0\delta t x(t)\hat{a}^{\dagger}\hat{a}]$. If $x(t)$ does not contain any randomness, the unitary induces a rotation in the optical phase space, $\hat{a}\rightarrow\exp[ig_0\delta t x(t)]\hat{a}$, which cannot change the minimal variance. If $x(t)$ contains classical randomness, the evolution is a convex mixture of rotations, with $\rho(t+\delta t)=\sum_j P_j \rho_j(t+\delta t)$, where $P_j$ is the probability for $x(t)$ to take value $x_j$, and $\rho_j(t+\delta t)$ is the evolution of the state based on $x_j$. The concavity of the variance for two probability distributions $P$ and $Q$, $\mathrm{Var}[\lambda P+(1-\lambda) Q]\geq\lambda\mathrm{Var}(P)+(1-\lambda)\mathrm{Var}(Q)$ with $\lambda\in[0,1]$, ensures that minimal variance cannot decrease, therefore squeezing for the mean field hybrid dynamics is not possible. This precluded squeezing from hybrid dynamics is because backaction from the oscillator to the field is not correctly captured.

Quantum and classical descriptions discussed in previous sections, in contrast, predict squeezing, as they capture the backaction via the correlation of amplitude and phase noise of the field mediated by the oscillator. In order to include the backaction, we will introduce a hybrid measurement model, where the oscillator evolves based on the instantaneous field intensity it detects and in turn collapses the field state continuously.

\section{Hybrid measurement model}

We present a description based on the hybrid quantum-classical model discussed in Ref.~\cite{diosi1988continuous}. This provides a way of modelling the quantum-classical interface, including the backaction. The classical oscillator effectively measures the intensity of the quantum field, thereby collapsing its wave function~\cite{jacobs2006straightforward,diosi1998coupling}. During an infinitesimal time step, the mechanical oscillator gains an infinitesimal amount of information about the instantaneous field intensity, and its position and momentum evolve consequently. In turn, the field evolves taking into account both the new position of the oscillator, and the fact that some classical information has been extracted by the oscillator (i.e. quantum wave function collapse). Note that, such measurement model is not directly applicable to experiments as there exists free parameter for continuous measurement rate (see discussions below). However, the idea of witnessing quantumness is that, the signature of quantumness must be incompatible with any theory where the subsystem we are interested in is dealt with classically. As will be demonstrated, the measurement model considered here will rule out optical squeezing as candidates of mechanical quantumness witness.

The equations of motion of the system are~\cite{jacobs2006straightforward,diosi1998coupling,wiseman2009quantum}
\begin{subequations}
\begin{align}
\begin{split}
d\rho&=ig_0xdt[\hat{a}^{\dagger}\hat{a},\rho]-\Gamma dt\Big[\hat{a}^{\dagger}\hat{a},[\hat{a}^{\dagger}\hat{a},\rho]\Big]\\
&~~~+\sqrt{2\Gamma}\Big(\{\hat{a}^{\dagger}\hat{a},\rho\}-2\mathrm{Tr}(\rho\hat{a}^{\dagger}\hat{a})\rho\Big)dW,
\end{split} \label{eq:measurement1} \\
dx&=\omega pdt, \label{eq:measurement2} \\
dp&=-\omega xdt+g_0\mathrm{Tr}(\rho\hat{a}^{\dagger}\hat{a})dt+\frac{g_0}{2\sqrt{2\Gamma}}dW, \label{eq:measurement3}
\end{align}
\end{subequations}
where $\Gamma$ characterises the rate at which classical information is gained by the oscillator, and $dW$ is the Wiener increment -- a zero mean Gaussian random variable with variance $dt$. Each term in Eqs.~\eqref{eq:measurement1}-\eqref{eq:measurement3} can be physically understood. The double commutator describes dephasing of the quantum field, and is stabilised around a Fock state; exactly which Fock state depends on one specific realisation of $dW$. The classical oscillator sees the effect of the field through the measurement of field intensity, which depends on both the average intensity and the classical fluctuations, thereby introducing noise correlations between the field and the oscillator.

Eqs.~\eqref{eq:measurement1}-\eqref{eq:measurement3} are solved numerically, and the resulting field variances as a function of time are shown in Fig.~\ref{fig:1}(c). Considering computational cost, we set $\alpha=2$ and $k=0.1$, keeping the product $\alpha k$ the same as that in the quantum description. This is a fair comparison as this product determines the maximum amount of squeezing (see Appendix). For the mechanical oscillator, we simulate both the case with $x(0)=p(0)=0$ (red line), and with $x(0)$ and $p(0)$ satisfying a Maxwell-Boltzmann distribution where their variances match the quantum zero-point-fluctuation (pink line). The continuous measurement strength $\Gamma$ is theoretically a free parameter, but here it is chosen as $\Gamma=0.01\omega$ such that both the optomechanical interaction and the continuous measurement contribute to the dynamics. Optical squeezing appears in both cases. The existence of initial thermal noise shows negligible influence on the field variance at the start of the evolution. In fact, the field is squeezed in every noise realisation, before taking the ensemble average.

Squeezing arises from the transient evolution from a coherent state to a specific Fock state, induced by the effective photon number measurement Eq.~\eqref{eq:measurement1}. The realization of $dW(t)$ continuously forces the field to evolve into a random Fock state. Once a specific Fock state is preferenced, the probability for the field to take other possible Fock states decreases, thus reducing the conditional uncertainty in the number basis. This number squeezed state can be well approximated by a quadrature squeezed state for a small amount of number squeezing. Note the squeezing only appears in the early stage of the time evolution, when the number squeezed $\sim$ quadrature squeezed approximation is valid. Eventually the state approaches a $dW(t)$-dependent Fock state, and averaging over all realizations of $dW(t)$ results in an incoherent mixture of Fock states~\cite{fox2006quantum}. Due to the non-unitarity at the field-oscillator interface, once information from the field flows into the oscillator, it cannot go back to the field. As a result, quantum revivals will not appear once the field variance increases to the stable value $2\alpha^2+1$, which corresponds to a Poissonian mixture of Fock state variance.

The squeezing is a result of the apparent weak measurement, not the unitary dynamics of the oscillator. As such, the amount of squeezing is always smaller than the case when, the field intensity is continuously measured by an apparatus without dynamics, shown as the blue line in Fig.~\ref{fig:1}(c) with $k=0$. This is because the unitary evolution of the oscillator induces a convex mixture of rotations of the optical field in phase space. If $k=0$, the maximum amount of squeezing is always at $\theta=0$ for a real $\alpha$. A comparison between squeezing in this hybrid measurement model and the quantum model is presented in the Appendix.


\section{Open cavity dynamics}

In the following we briefly discuss how photons exit the cavity to be detected. Note that opening the cavity to vacuum fluctuations will blur witnesses of mechanical quantumness in all cases. For our parameters intracavity squeezing appears over the time scale of the mechanical period $\tau$. Using the cavity input-output relations~\cite{collett1984squeezing}, and considering a highly dissipative cavity, the output field is dominated by the intracavity squeezed field. The green dashed line in Fig.~\ref{fig:1}(c) simulates the intracavity field variance in the presence of photon dissipation with rate $\kappa=\omega$, which is required to observe squeezing on this time scale. The cavity decay is simulated with the standard quantum optical Lindblad term in the master equation~\cite{gardiner2004quantum}. The squeezing in the output field may be observed using time-binned homodyne detection~\cite{smithey1993measurement}, and choosing the temporal profile of the local oscillator to capture only the squeezed part of the outgoing field~\cite{opatrny1997homodyne}.

For comparison, we plot the field variance predicted by the quantum description in the presence of dissipation $\kappa=0.3\omega$ (Fig.~\ref{fig:1}(c), brown dotted line). We note that such cavity decay rates --- required to observe short-time squeezing in the output field --- destroy long term temporal correlations in the optical field, precluding the observation of optical revivals. Cavity loss has a stronger effect as the squeezing is reduced to a level similar to that of the hybrid measurement model with a larger dissipation rate. Note that maximal squeezing appears later for the quantum description.

\section{Conclusion}

We discuss the plausibility of using optical squeezing to demonstrate the necessity of quantisation of the inaccessible mechanical oscillator in an optomechanical system. Although we model the optomechanical system as a nonlinear closed system which maximises the possibility of transferring mechanical quantumness into the field mode, we still find that optical squeezing does not in general indicate mechanical quantumness. Treating the oscillator classically, we are able to recover optical squeezing as long as the backaction onto the field is captured. Recurrence of optical squeezing cannot be reproduced by the alternative descriptions discussed here, thus being a potential witness of mechanical quantumness. However, its observation requires at least single-photon blockade condition, which is impractical given current technologies.

\acknowledgements  

MSK and KK acknowledge the Leverhulme Trust [Project RPG-2014-055] and the Royal Society. FA and MSK acknowledge the Marie Curie Actions of the EU's 7$^{\mbox{th}}$ Framework Programme under REA [grant number 317232]. YM is supported by the EPSRC Centre for Doctoral Training on Controlled Quantum Dynamics at Imperial College London (EP/L016524/1) and funded by the Imperial College PresidentÕs PhD Scholarship.

\newpage

\appendix

\section{Analytical expressions of variance in the quantum, classical and hybrid mean field descriptions}\label{app:1}
The full expression of the field variance in the quantum description is given by
\begin{equation}
\begin{aligned}
&{\rm Var}_\theta^{(q)}(t)=\langle\hat{\sf X}_{\theta}^2(t)\rangle-\langle\hat{\sf X}_{\theta}(t)\rangle^2=\\
&2\alpha^2e^{-\alpha^2\big(1-\cos2A(t)\big)}e^{-2B(t)}\cos\big(2A(t)+\alpha^2\sin2A(t)-2\theta\big)\\
&-2\alpha^2e^{-2\alpha^2\big(1-\cos A(t)\big)}e^{-B(t)}\cos\big(A(t)+2\alpha^2\sin A(t)-2\theta\big)\\
&+2\alpha^2\Big(1-e^{-2\alpha^2\big(1-\cos A(t)\big)}e^{-B(t)}\Big)+1.
\end{aligned}
\label{eq:quantum_variance}
\end{equation}
Here we consider the initial state of the intracavity field to be a coherent state $|\alpha\rangle_{\rm L}$, and the mechanical oscillator to be a thermal state $\hat{\rho}_{\rm th}=\sum_{n=0}^{\infty}p_n|n\rangle\langle n|$ at temperature $T$, where $p_n=\tilde{n}_{\rm th}^n/(\tilde{n}_{\rm th}+1)^{n+1}$ with $\tilde{n}_{\rm th}=1/(\exp(\hbar\omega/k_BT)-1)$, and $\ket{n}$ a Fock state. The functions are defined as $A(t)=2k^2(\omega t-\sin\omega t)$ and $B(t)=2k^2(2\tilde{n}_{\mathrm{th}}+1)(1-\cos\omega t)$.

To express the full expression of the field variance as a function of time in the classical description, we first define several functions to simplify the notation:
\begin{subequations}
\begin{align}
&d_1(k,\omega,t)=2A^2(t)/\big(1+A^2(t)\big),\\
&d_2(k,\omega,t)=2A^2(t)/\big(4+A^2(t)\big),\\
&d_3(k,\omega,t)=16/\big(4+A^2(t)\big)^2,\\
&c_1(k,\omega,t)=\big(1-3A^2(t)\big)/\big(1+A^2(t)\big)^3,\\
&c_2(k,\omega,t)=\big(256-384A^2(t)+16A^4(t)\big)/\big(4+A^2(t)\big)^4,\\
&s_1(k,\omega,t)=\big(3A(t)-A^3(t)\big)/\big(1+A^2(t)\big)^3,\\
&s_2(k,\omega,t)=\big(512A(t)-128A^3(t)\big)/\big(4+A^2(t)\big)^4,\\
&\phi_1(k,\omega,t)=2A(t)/\big(1+A^2(t)\big),\\
&\phi_2(k,\omega,t)=4A(t)/\big(4+A^2(t)\big).
\end{align}
\end{subequations}
The variance in the classical description is then
\begin{widetext}
\begin{equation}
\begin{aligned}
&{\rm Var}_{\theta}^{(c)}(t)\\
&=2\alpha^2e^{-\alpha^2d_1(k,\omega,t)}e^{-2C(t)}\Big(c_1(k,\omega,t)\cos\big(\alpha^2\phi_1(k,\omega,t)-2\theta\big)-s_1(k,\omega,t)\sin\big(\alpha^2\phi_1(k,\omega,t)-2\theta\big)\Big)\\
&-2\alpha^2e^{-2\alpha^2d_2(k,\omega,t)}e^{-C(t)}\Big(c_2(k,\omega,t)\cos\big(2\alpha^2\phi_2(k,\omega,t)-2\theta\big)-s_2(k,\omega,t)\sin\big(2\alpha^2\phi_2(k,\omega,t)-2\theta\big)\Big)\\
&+2\alpha^2\Big(1-d_3(k,\omega,t)e^{-2\alpha^2d_2(k,\omega,t)}e^{-C(t)}\Big)+1.
\end{aligned}
\label{eq:classical_variance}
\end{equation}
\end{widetext}
Here we define $C(t)=4n_{\rm th}k^2(1-\cos\omega t)$ with $n_{\rm th}=k_BT/(\hbar\omega_{\rm M})$ as the classical mean thermal excitation number. Here we have also assumed that $\alpha$ is real.

The existence of quantum revivals in the quantum description can be understood in the following way. The overall behavior of the quantum variance is controlled by the exponentials $\exp[-\alpha^2(1-\cos2A(t))]$ and $\exp[-2\alpha^2(1-\cos A(t))]$. Considering that $\alpha^2\gg1$, the two exponentials are nonzero only when the cosine terms are close to $1$. To be specific, when $\omega t\approx N\pi/2k^2$ where $N$ is an odd integer, $\exp[-\alpha^2(1-\cos2A(t))]\approx1$ but $\exp[-2\alpha^2(1-\cos A(t))]\approx0$. The quantum variance is approximated as 
\begin{equation}
{\rm Var}_{\theta}^{(q)}(t)\approx2\alpha^2\cos\Big(2A(t)+\alpha^2\sin2A(t)-2\theta\Big)+2\alpha^2+1,
\label{eq:zero}
\end{equation}
which represents the first quantum revival. When $\omega t\approx N\pi/k^2$ where $N$ is an integer, $\exp[-\alpha^2(1-\cos2A(t))]\approx1$ and $\exp[-2\alpha^2(1-\cos A(t))]\approx1$. In this case the quantum variance is approximated as
\begin{equation}
\begin{aligned}
&{\rm Var}_{\theta}^{(q)}(t)\\
\approx&2\alpha^2\Big(\cos\big(2A(t)+\alpha^2\sin2A(t)-2\theta\big)-\cos\big(A(t)+\\
&2\alpha^2\sin A(t)-2\theta\big)\Big)+1,
\end{aligned}
\label{eq:squeeze}
\end{equation}
which, due to the lack of the constant contribution $2\alpha^2$, shows an overall reduction as in the second quantum revival. Note that, for simplicity, we have made the approximations $\exp[-4k^2(1-\cos\omega t)]\approx1$ and $\exp[-2k^2(1-\cos\omega t)]\approx1$, which is reasonable as $k^2\ll1$.

In contrast, the overall behavior of the classical variance is controlled by $\exp[-\alpha^2d_1(k,\omega,t)]$ and $\exp[-2\alpha^2d_2(k,\omega,t)]$. But $d_1$ and $d_2$ are close to zero only at the beginning of the interaction. Once $\exp[-\alpha^2d_1(k,\omega,t)]$ and $\exp[-2\alpha^2d_2(k,\omega,t)]$ decay to zero, the classical variance stays at $2\alpha^2+1$ without further oscillations.

We have considered three hybrid descriptions based on mean field approximation in the main text. If the oscillator sees only the mean intensity of the field, the variance is
\begin{equation}
\begin{aligned}
&{\rm Var}^{(sc1)}_{\theta}(t)\\
&=1+2\alpha^2\big(1-e^{-C(t)}\big)\Big(1-\cos\big(2\alpha^2A(t)-2\theta\big)e^{-C(t)}\Big).
\end{aligned}
\label{eq:semiclassical_variance}
\end{equation}
If the oscillator sees the energy quanta of the field with field intensity given by a random number following Poisson distribution, the variance becomes
\begin{equation}
\begin{aligned}
&\mathrm{Var}^{(sc2)}_{\theta}(t)\\
&=2\alpha^2e^{-\alpha^2\big(1-\cos 2A(t)\big)}e^{-2C(t)}\cos\big(\alpha^2\sin2A(t)-2\theta\big)\\
&~~~-2\alpha^2e^{-2\alpha^2\big(1-\cos A(t)\big)}e^{-C(t)}\cos\big(2\alpha^2\sin A(t)-2\theta\big)\\
&~~~+2\alpha^2\Big(1-e^{-C(t)}e^{-2\alpha^2\big(1-\cos A(t)\big)}\Big)+1.
\end{aligned}
\label{eq:poisson}
\end{equation}
And if the Poisson distribution is approximated by a Gaussian distribution, the variance turns out to be
\begin{equation}
\begin{aligned}
{\rm Var}^{(sc3)}_{\theta}(t)&=1+2\alpha^2\big(1-e^{-C(t)}e^{-\alpha^2A^2(t)}\big)\\
&\times\Big(1-\cos\big(2\alpha^2A(t)-2\theta\big)e^{-C(t)}e^{-\alpha^2A^2(t)}\Big).
\end{aligned}
\label{eq:semiclassical_variance2}
\end{equation}

\section{Thermal initial state and thermal bath for the oscillator}\label{app:2}

In the main text we focus on the case where the mechanical oscillator starts from a vacuum state, and it is not subject to thermal noise throughout the evolution. In this section we will analyze the effect of both factors, respectively.

\begin{figure}[t]
\centering
\includegraphics[width=0.45\textwidth]{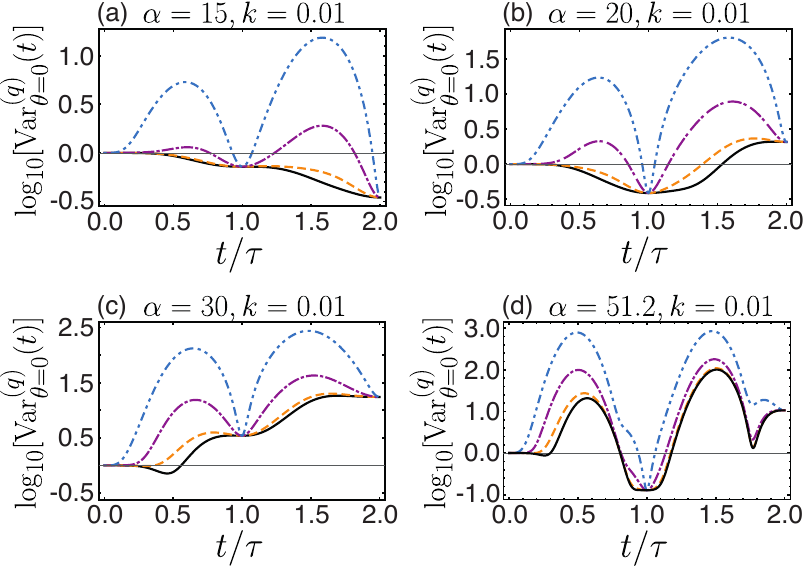}
\caption{Variance of the cavity field as a function of time for different combinations of $\alpha$ and $k$ in the quantum picture. Black solid line corresponds to the case where the oscillator starts from the vacuum state. Green dotted line is for the Kerr nonlinear medium that is equivalent to the optomechanical model at the end of each mechanical period. Orange dashed line is for $\tilde{n}_{\rm th}=1$ ($T=2.1\ {\rm mK}$), purple dot-dashed line is for $\tilde{n}_{\rm th}=10$ ($T=15.1\ {\rm mK}$), and blue dot-dot-dashed line is for $\tilde{n}_{\rm th}=100$ ($T=144.8\ {\rm mK}$), where $\omega=2\pi\times30\ {\rm MHz}$ is assumed.
}\label{fig:thermal}
\end{figure}

The thermal initial state is well captured by the parameter $\tilde{n}_{\mathrm{th}}$ in Eq.~\eqref{eq:quantum_variance}. A vacuum state corresponds to $\tilde{n}_{\mathrm{th}}=0$. The initial thermal excitations bring the field variance closer to the constant value $2\alpha^2+1$. However, at the end of each mechanical period, the oscillator and the field decouples. At those times the field does not see the thermal noise in the oscillator any more. Fig.~\ref{fig:thermal} shows four examples of how the field variance changes with time, with different thermal initial states of the oscillator. The periodical recombinations of variances with different initial thermal phonons are clear. The variances at the end of each mechanical period coincide with the case of the equivalent model of Kerr medium as well. Note here we set $\theta=0$ for simplicity.

\begin{figure}[t!]
\centering
\includegraphics[width=0.45\textwidth]{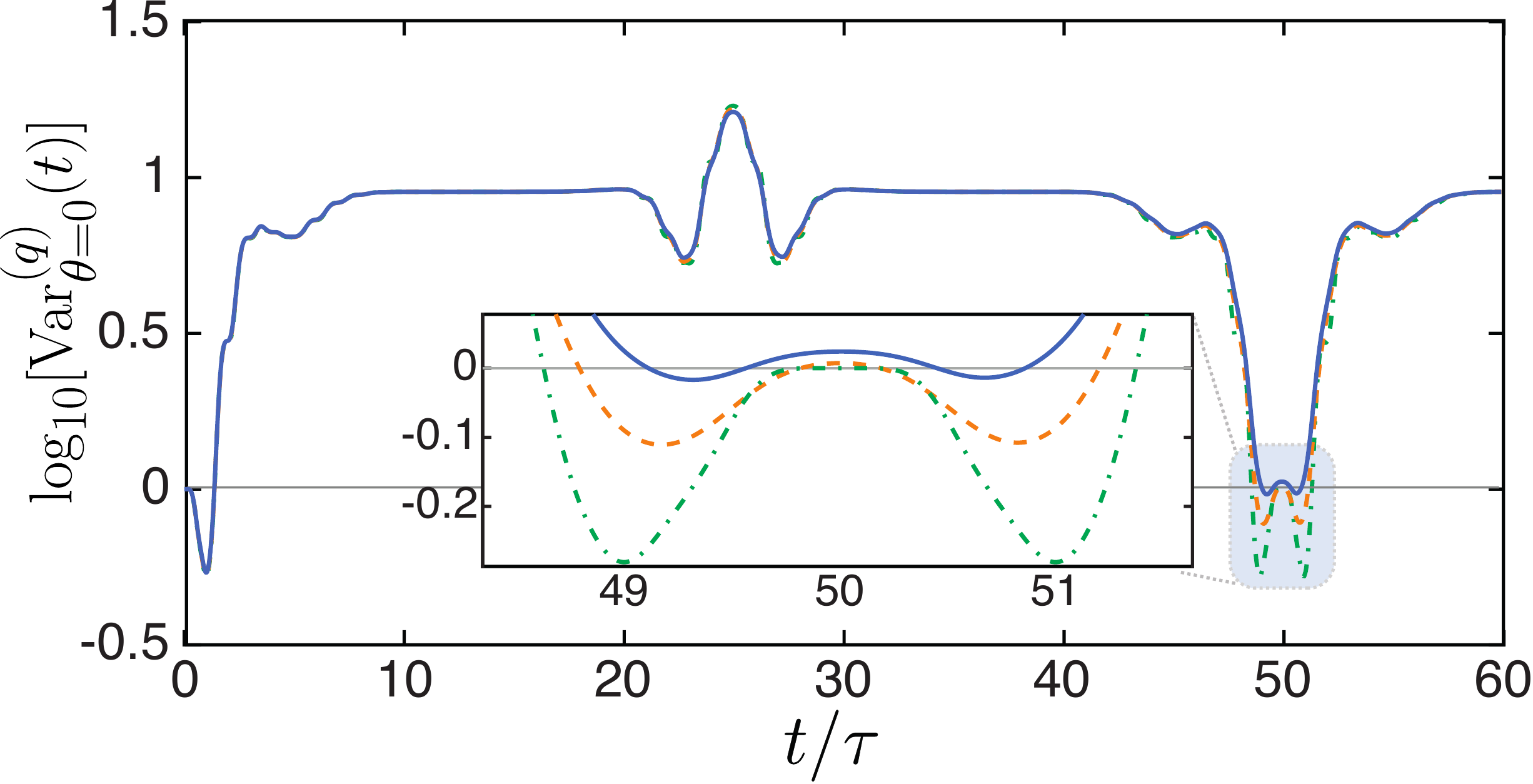}
\caption{Variance of the cavity field in the quantum picture as a function of time when the mechanical oscillator is in contact with a thermal bath. $\alpha=2$, $k=0.1$. Other parameters are: $\gamma/\omega=0$, $\tilde{n}=0$ (green dot-dashed line), $\gamma/\omega=0.01$, $\tilde{n}=0$ (orange dashed line), $\gamma/\omega=0.01$, $\tilde{n}=0.5$ (blue solid line).}
\label{fig:damp}
\end{figure}

Continuous contact of the oscillator with a thermal bath has a different impact on the field variance. In the weak coupling regime, the dynamics is governed by the master equation~\cite{bose1997preparation,nunnenkamp2011single}
\begin{equation}
\begin{aligned}
&\dot{\rho}(t)=\\
&-i[\hat{H},\rho(t)]/\hbar+\gamma(\tilde{n}+1)\Big(\hat{b}\rho(t)\hat{b}^{\dagger}-\frac{1}{2}\hat{b}^{\dagger}\hat{b}\rho(t)-\frac{1}{2}\rho(t)\hat{b}^{\dagger}\hat{b}\Big)\\
&+\gamma\tilde{n}\Big(\hat{b}^{\dagger}\rho(t)\hat{b}-\frac{1}{2}\hat{b}\hat{b}^{\dagger}\rho(t)-\frac{1}{2}\rho(t)\hat{b}\hat{b}^{\dagger}\Big),
\end{aligned}
\label{eq:master}
\end{equation}
where $\hat{H}$ is the Hamiltonian
\begin{equation}
\hat{H}/\hbar=\omega\hat{b}^{\dagger}\hat{b}-\frac{g_0}{\sqrt{2}}\hat{a}^{\dagger}\hat{a}(\hat{b}^{\dagger}+\hat{b}),
\label{eq:hamiltonian}
\end{equation}
$\tilde{n}$ is the mean phonon number of the bath, and $\gamma$ characterizes the mechanical decay rate. Fig.~\ref{fig:damp} shows the numerical result of how the field variance evolves with $\alpha=2$ and $k=0.1$ considering computational cost. We assume the oscillator starts from a vacuum state for simplicity, and set $\theta=0$ for the variance. The variance gradually gets closer to the constant value $2\alpha^2+1$. Clearly as zoomed in, the amplitude of quantum revivals is reduced.

\section{Dependence of field variance on parameters $\alpha$ and $k$}\label{app:3}

The coherent state amplitude $\alpha$ and interaction strength $k$ both affect how the field variance evolves with time. In this section we take the quantum description as an example to analyze those effects.

\begin{figure}[t!]
\centering
\includegraphics[width=0.47\textwidth]{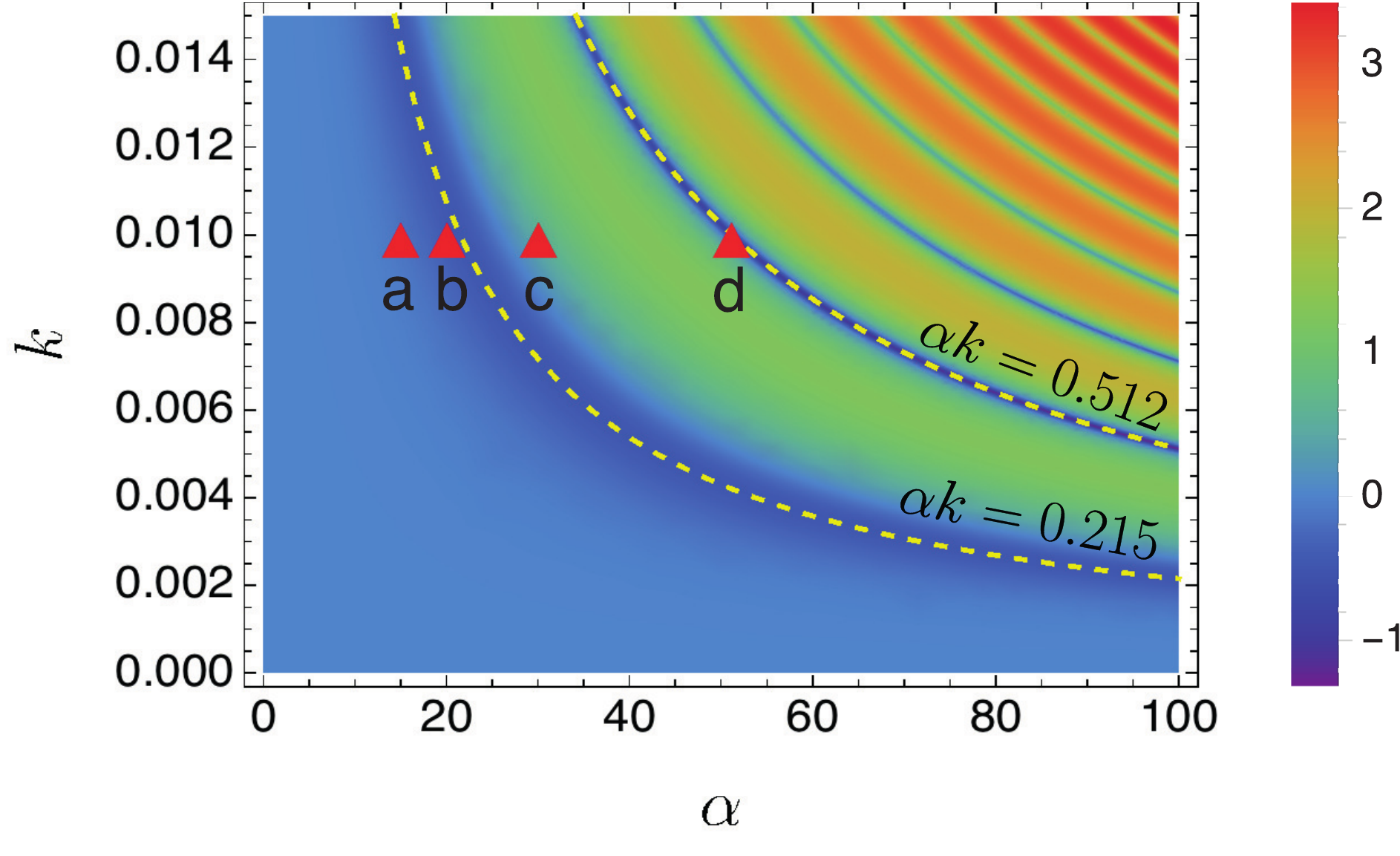}
\caption{Variance of the cavity field as a function of the coherent state amplitude $\alpha$ and the rescaled coupling strength $k$ in the quantum picture. Color shows the value of $\log_{10}[{\rm Var}_{\theta=0}^{(q)}(t=\tau)]$. Red triangles correspond to the four cases in Fig.~\ref{fig:thermal}.}
\label{fig:colour_variance1}
\end{figure}

As seen from Fig.~\ref{fig:thermal}, at the end of each mechanical period, the time derivative of field variance is close to zero, which means the variance does not change rapidly in time. Also the initial thermal phonon excitation does not affect the variance at those times. Considering that the side band resolution $\omega/\kappa$ is usually smaller than $1$ or on the order of $1$, we choose the variance at $t=\tau$ to represent the amount of squeezing. Usually $\alpha k<1$ is satisfied. Once we Taylor expand Eq.~\eqref{eq:quantum_variance} around $k=0$, we can see that the product $\alpha k$ directly decides the variance. This is clearly shown in Fig.~\ref{fig:colour_variance1}. Especially, the two contours marked in yellow indicate squeezing, and for those two contours $\theta=0$ is very close to the angle corresponding to the minimum variance.

$k$ itself decides quantum revivals times. As discussed in the main text, revival times are proportional to $1/k^2$.

\section{Derivation of the equations of motion in the fully classical picture}\label{app:4}

The physical picture in the classical description is very clear. The mechanical oscillator provides a moving instantaneous boundary to the field, while the field exerts a radiation pressure force on the mechanical oscillator depending on the field intensity~\cite{law1995interaction}. As pointed out in Ref.~\cite{armata2016quantum}, the field acquires a phase depending on the distance it travels inside the cavity, which is determined by the position of the mechanical oscillator. However, the variance of the field is non-trivially related to both amplitude and phase of the field, which requires a careful analysis. Our starting point is the classical Hamiltonian describing the dynamics of a one-dimensional electromagnetic field in a cavity with a movable boundary. Especially, the degrees of freedom of the boundary must be included in the overall system's dynamics. The Hamiltonian of the system is found to be~\cite{law1995interaction}
\begin{equation}
\tilde{H}=\frac{p_{\rm M}^2}{2M}+\frac{1}{2}M\omega^2x_{\rm M}^2+\frac{1}{2}(P_{\rm L}^2+\Omega^2Q_{\rm L}^2)-\frac{x_{\rm M}}{L}\Omega^2Q_{\rm L}^2,
\end{equation}
where $\{x_{\rm M},p_{\rm M}\}$ and $\{Q_{\rm L},P_{\rm L}\}$ are pairs of canonical variables for the mechanical oscillator and the field, respectively. To simplify the notations, we rescale the canonical variables as $x=\sqrt{M\omega}x_{\rm M}$, $p=p_{\rm M}/\sqrt{M\omega}$, $\tilde{x}_{\rm L}=\sqrt{\Omega}Q_{\rm L}$ and $\tilde{p}_{\rm L}=P_{\rm L}/\sqrt{\Omega}$. Note that the rescaling keeps the Poisson bracket invariant, so $\{x,p,\tilde{x}_{\rm L},\tilde{p}_{\rm L}\}$ act as canonical variables as well~\cite{hand1998analytical}. Now the Hamiltonian becomes
\begin{equation}
\tilde{H}=\frac{1}{2}\omega(x^2+p^2)+\frac{1}{2}\Omega(\tilde{x}_{\rm L}^2+\tilde{p}_{\rm L}^2)-g_0x\tilde{x}_{\rm L}^2.
\end{equation}
We perform a canonical transformation which maps $\{\tilde{x}_{\rm L},\tilde{p}_{\rm L}\}$ into another pair of canonical variables $\{x_{\rm L},p_{\rm L}\}$ in the frame rotating with frequency $\Omega$, namely
\begin{equation}
\begin{aligned}
x_{\rm L}&=\tilde{x}_{\rm L}\cos\Omega t-\tilde{p}_{\rm L}\sin\Omega t,\\
p_{\rm L}&=\tilde{x}_{\rm L}\sin\Omega t+\tilde{p}_{\rm L}\cos\Omega t.
\end{aligned}
\label{eq:transform}
\end{equation}
The equations of motion are then found to be
\begin{equation}
\begin{aligned}
\frac{dp(t)}{dt}=&-\omega x(t)+g_0\Big(x_{\rm L}^2(t)\cos^2\Omega t+p_{\rm L}^2(t)\sin^2\Omega t\\
&+x_{\rm L}(t)p_{\rm L}(t)\sin2\Omega t\Big),\\
\frac{dx(t)}{dt}=&\omega p(t),\\
\frac{dp_{\rm L}(t)}{dt}=&g_0x(t)\Big(2x_{\rm L}(t)\cos^2\Omega t+p_{\rm L}(t)\sin2\Omega t\Big),\\
\frac{dx_{\rm L}(t)}{dt}=&-g_0x(t)\Big(x_{\rm L}(t)\sin2\Omega t+2p_{\rm L}(t)\sin^2\Omega t\Big).
\end{aligned}
\label{eq:EOMclassical2}
\end{equation}
Note that in Eq.~\eqref{eq:EOMclassical2} the first order time derivatives of the canonical variables do not depend on $\Omega x(t)$, $\Omega p(t)$, $\Omega x_{\rm L}(t)$ or $\Omega p_{\rm L}(t)$, which means that the fast oscillation of the light is already removed by the transformation Eq.~\eqref{eq:transform}. The canonical variables evolve on a time scale much slower than the time scale set by the field frequency $\Omega$ under usual optomechanical parameters. As a result, we can approximate the equations of motion by taking the time average over the period of the field, which leaves everything unchanged except those related to $\Omega t$, namely, $\cos^2\Omega t\rightarrow1/2$, $\sin^2\Omega t\rightarrow1/2$ and $\sin\Omega t\cos\Omega t\rightarrow0$. The effective equations of motion then become
\begin{equation}
\begin{aligned}
\frac{dp(t)}{dt}=&-\omega x(t)+\frac{1}{2}g_0\Big(x_{\rm L}^2(t)+p_{\rm L}^2(t)\Big),\\
\frac{dx(t)}{dt}=&\omega p(t),\\
\frac{dp_{\rm L}(t)}{dt}=&g_0x(t)x_{\rm L}(t),\\
\frac{dx_{\rm L}(t)}{dt}=&-g_0x(t)p_{\rm L}(t).
\end{aligned}
\label{eq:EOMclassical3}
\end{equation}
It is straightforward to check that Eq.~\eqref{eq:EOMclassical3} corresponds to the Hamiltonian
\begin{equation}
H=\omega |\beta|^2-\frac{g_0}{\sqrt{2}}|\alpha_{\rm L}|^2(\beta^{*}+\beta),
\label{eq:approx_Hamiltonian}
\end{equation}
once we define a complex variable describing the field as $\alpha_{\rm L}=(x_{\rm L}+ip_{\rm L})/\sqrt{2}$ and a complex variable describing the mechanical oscillator as $\beta=(x+ip)/\sqrt{2}$. Direct comparison with Eq.~\eqref{eq:hamiltonian} implies that, $\alpha_{\rm L}(\beta)$ is the classical version of $\hat{a}(\hat{b})$. 

The transformation Eq.~\eqref{eq:transform} together with the average over fast oscillation is essentially the classical counterpart of the rotating wave approximation which is necessary in the derivation of the standard optomechanical Hamiltonian~\eqref{eq:hamiltonian}, where, in fact, the terms proportional to $a^2$ and $a^{\dagger2}$ are neglected~\cite{law1995interaction,sala2018exploring}. It enables us to solve the equations of motion, Eq.~\eqref{eq:EOMclassical3}, analytically. It is straightforward to check that $\{|\alpha_{\rm L}|^2,H\}=0$, where $\{\cdot,\cdot\}$ represents the Poisson bracket. As a result, $|\alpha_{\rm L}|^2$ is a conserved quantity during the time evolution, as in the quantum case. We can thus directly replace $x_{\rm L}^2(t)+p_{\rm L}^2(t)$ with $x_{\rm L}^2(0)+p_{\rm L}^2(0)$ in Eq.~\eqref{eq:EOMclassical3}. The solutions of Eq.~\eqref{eq:EOMclassical3} are then given by
\begin{equation}
\begin{aligned}
x(t)=&x(0)\cos\omega t+p(0)\sin\omega t+\frac{g_0}{\omega}|\alpha_{\rm L}(0)|^2(1-\cos\omega t),\\
p(t)=&-x(0)\sin\omega t+p(0)\cos\omega t+\frac{g_0}{\omega}|\alpha_{\rm L}(0)|^2\sin\omega t,\\
\alpha_{\rm L}(t)=&\alpha_{\rm L}(0)e^{ig_0\int_0^tx(\tau)d\tau}\\
=&\alpha_{\rm L}(0)e^{iA(t)|\alpha_{\rm L}(0)|^2}e^{i\frac{g_0}{\omega}\big(\sin\omega tx(0)+(1-\cos\omega t)p(0)\big)},
\end{aligned}
\label{eq:classicalEOM}
\end{equation}
where $x(t)$ ($p(t)$) represents the position (momentum) of the mechanical oscillator, $\alpha_{\mathrm{L}}(t)$ represents the amplitude of the field.

\section{comparing the squeezing in quantum and hybrid measurement model}

\begin{figure}[h]
\centering
\includegraphics[width=0.45\textwidth]{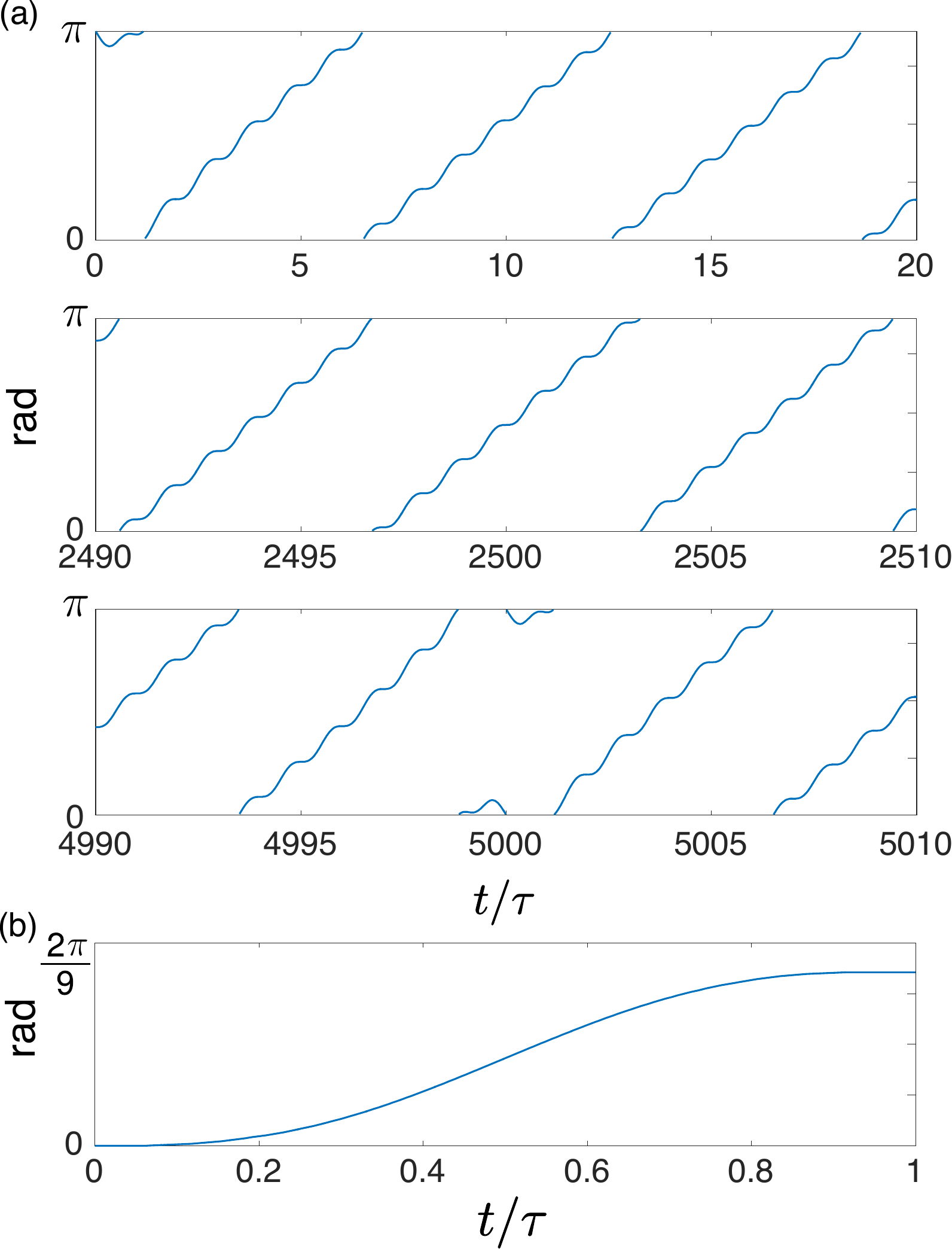}
\caption{Angle $\theta$ that minimises the variance $\mathrm{Var}_{\theta}(t)$ as a function of time. (a) The quantum mechanical description. The sharp jumps from $180^{\circ}$ to $0^{\circ}$ are due to the fact that they correspond to the same value of $\mathrm{Var}_{\theta}(t)$. (b) The hybrid continuous measurement model.
}\label{fig:phase}
\end{figure}

\begin{figure}[h]
\centering
\includegraphics[width=0.45\textwidth]{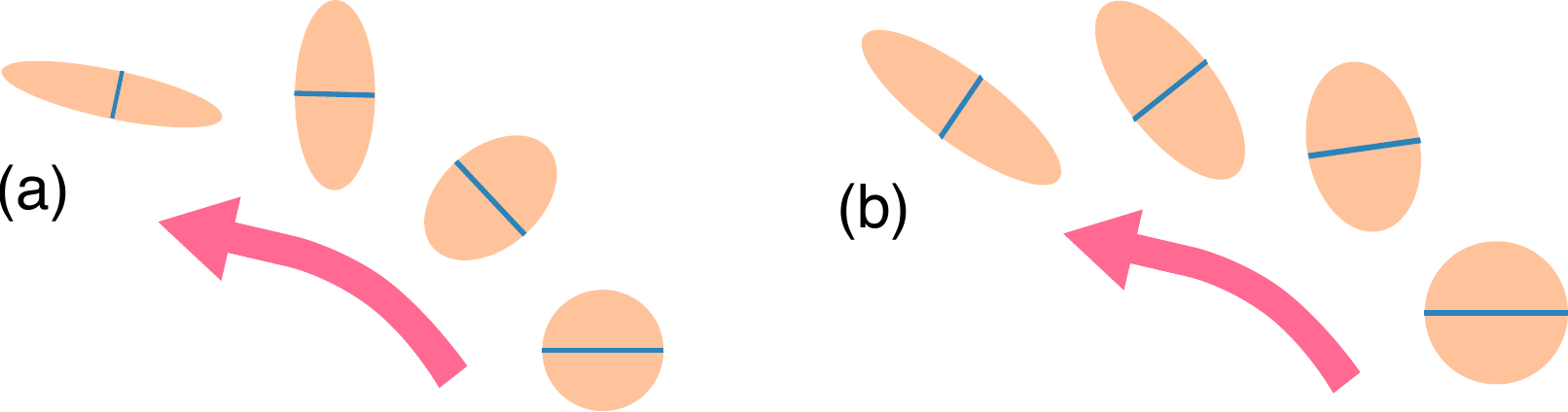}
\caption{Illustration of the time evolution of uncertainty circle. Blue bar indicates the quadrature that corresponds to the minimum variance. (a) The quantum mechanical description. (b) The hybrid continuous measurement model.
}\label{fig:illu}
\end{figure}

We plot the time evolution of phase angle $\theta$ which minimises the variance $\mathrm{Var}_{\theta}(t)$, in the quantum mechanical description (Fig.~\ref{fig:phase}(a)) and hybrid continuous measurement model (Fig.~\ref{fig:phase}(b)). As discussed in the main text, the origins of squeezing are different. In the quantum mechanical description, the squeezing originates from the intensity-dependent phase shift. The uncertainty circle is stretched so that points further away from the centre of the phase space have a larger displacement. It causes $\theta$ to decrease from $180^{\circ}$. However, the mechanical oscillator dynamics induces a global rotation of the uncertainty circle around the centre of the phase space. It brings the angle $\theta$ back to $180^{\circ}$ (equivalently $0^{\circ}$) and then $\theta$ increases from $0^{\circ}$. The process results in the dips in Fig.~\ref{fig:phase}(a) around $t/\tau=0$ and $t/\tau=5000$, and is illustrated in Fig.~\ref{fig:illu}(a). For the hybrid measurement model, however, squeezing is a result of the transition from coherent state to a Fock state induced by the measurement of field intensity. Without unitary dynamics of the oscillator, $\theta$ is always zero. When the unitary dynamics is included, the uncertainty circle goes through global rotation, which causes angle $\theta$ to increase from $0^{\circ}$. The process is illustrated in Fig.~\ref{fig:illu}(b).

\newpage

\bibliographystyle{apsrev4-1} 
\bibliography{MyReferenceShort} 

\begin{thebibliography}{67}%
\makeatletter
\providecommand \@ifxundefined [1]{%
 \@ifx{#1\undefined}
}%
\providecommand \@ifnum [1]{%
 \ifnum #1\expandafter \@firstoftwo
 \else \expandafter \@secondoftwo
 \fi
}%
\providecommand \@ifx [1]{%
 \ifx #1\expandafter \@firstoftwo
 \else \expandafter \@secondoftwo
 \fi
}%
\providecommand \natexlab [1]{#1}%
\providecommand \enquote  [1]{``#1''}%
\providecommand \bibnamefont  [1]{#1}%
\providecommand \bibfnamefont [1]{#1}%
\providecommand \citenamefont [1]{#1}%
\providecommand \href@noop [0]{\@secondoftwo}%
\providecommand \href [0]{\begingroup \@sanitize@url \@href}%
\providecommand \@href[1]{\@@startlink{#1}\@@href}%
\providecommand \@@href[1]{\endgroup#1\@@endlink}%
\providecommand \@sanitize@url [0]{\catcode `\\12\catcode `\$12\catcode
  `\&12\catcode `\#12\catcode `\^12\catcode `\_12\catcode `\%12\relax}%
\providecommand \@@startlink[1]{}%
\providecommand \@@endlink[0]{}%
\providecommand \url  [0]{\begingroup\@sanitize@url \@url }%
\providecommand \@url [1]{\endgroup\@href {#1}{\urlprefix }}%
\providecommand \urlprefix  [0]{URL }%
\providecommand \Eprint [0]{\href }%
\providecommand \doibase [0]{http://dx.doi.org/}%
\providecommand \selectlanguage [0]{\@gobble}%
\providecommand \bibinfo  [0]{\@secondoftwo}%
\providecommand \bibfield  [0]{\@secondoftwo}%
\providecommand \translation [1]{[#1]}%
\providecommand \BibitemOpen [0]{}%
\providecommand \bibitemStop [0]{}%
\providecommand \bibitemNoStop [0]{.\EOS\space}%
\providecommand \EOS [0]{\spacefactor3000\relax}%
\providecommand \BibitemShut  [1]{\csname bibitem#1\endcsname}%
\let\auto@bib@innerbib\@empty
\bibitem [{\citenamefont {Kot}\ \emph {et~al.}(2012)\citenamefont {Kot},
  \citenamefont {Gr{\o}nbech-Jensen}, \citenamefont {Nielsen}, \citenamefont
  {Neergaard-Nielsen}, \citenamefont {Polzik},\ and\ \citenamefont
  {S{\o}rensen}}]{kot2012breakdown}%
  \BibitemOpen
  \bibfield  {author} {\bibinfo {author} {\bibfnamefont {E.}~\bibnamefont
  {Kot}}, \bibinfo {author} {\bibfnamefont {N.}~\bibnamefont
  {Gr{\o}nbech-Jensen}}, \bibinfo {author} {\bibfnamefont {B.~M.}\ \bibnamefont
  {Nielsen}}, \bibinfo {author} {\bibfnamefont {J.~S.}\ \bibnamefont
  {Neergaard-Nielsen}}, \bibinfo {author} {\bibfnamefont {E.~S.}\ \bibnamefont
  {Polzik}}, \ and\ \bibinfo {author} {\bibfnamefont {A.~S.}\ \bibnamefont
  {S{\o}rensen}},\ }\href@noop {} {\bibfield  {journal} {\bibinfo  {journal}
  {Phys. Rev. Lett.}\ }\textbf {\bibinfo {volume} {108}},\ \bibinfo {pages}
  {233601} (\bibinfo {year} {2012})}\BibitemShut {NoStop}%
\bibitem [{\citenamefont {Bell}(1964)}]{bell1964einstein}%
  \BibitemOpen
  \bibfield  {author} {\bibinfo {author} {\bibfnamefont {J.~S.}\ \bibnamefont
  {Bell}},\ }\href@noop {} {\bibfield  {journal} {\bibinfo  {journal} {Physics
  Physique Fizika}\ }\textbf {\bibinfo {volume} {1}},\ \bibinfo {pages} {195}
  (\bibinfo {year} {1964})}\BibitemShut {NoStop}%
\bibitem [{\citenamefont {Vivoli}\ \emph {et~al.}(2016)\citenamefont {Vivoli},
  \citenamefont {Barnea}, \citenamefont {Galland},\ and\ \citenamefont
  {Sangouard}}]{vivoli2016proposal}%
  \BibitemOpen
  \bibfield  {author} {\bibinfo {author} {\bibfnamefont {V.~C.}\ \bibnamefont
  {Vivoli}}, \bibinfo {author} {\bibfnamefont {T.}~\bibnamefont {Barnea}},
  \bibinfo {author} {\bibfnamefont {C.}~\bibnamefont {Galland}}, \ and\
  \bibinfo {author} {\bibfnamefont {N.}~\bibnamefont {Sangouard}},\ }\href@noop
  {} {\bibfield  {journal} {\bibinfo  {journal} {Phys. Rev. Lett.}\ }\textbf
  {\bibinfo {volume} {116}},\ \bibinfo {pages} {070405} (\bibinfo {year}
  {2016})}\BibitemShut {NoStop}%
\bibitem [{\citenamefont {Marinkovic}\ \emph {et~al.}(2018)\citenamefont
  {Marinkovic}, \citenamefont {Wallucks}, \citenamefont {Riedinger},
  \citenamefont {Hong}, \citenamefont {Aspelmeyer},\ and\ \citenamefont
  {Gr{\"o}blacher}}]{marinkovic2018optomechanical}%
  \BibitemOpen
  \bibfield  {author} {\bibinfo {author} {\bibfnamefont {I.}~\bibnamefont
  {Marinkovic}}, \bibinfo {author} {\bibfnamefont {A.}~\bibnamefont
  {Wallucks}}, \bibinfo {author} {\bibfnamefont {R.}~\bibnamefont {Riedinger}},
  \bibinfo {author} {\bibfnamefont {S.}~\bibnamefont {Hong}}, \bibinfo {author}
  {\bibfnamefont {M.}~\bibnamefont {Aspelmeyer}}, \ and\ \bibinfo {author}
  {\bibfnamefont {S.}~\bibnamefont {Gr{\"o}blacher}},\ }\href@noop {}
  {\bibfield  {journal} {\bibinfo  {journal} {arXiv preprint arXiv:1806.10615}\
  } (\bibinfo {year} {2018})}\BibitemShut {NoStop}%
\bibitem [{\citenamefont {Glauber}(1963)}]{glauber1963coherent}%
  \BibitemOpen
  \bibfield  {author} {\bibinfo {author} {\bibfnamefont {R.~J.}\ \bibnamefont
  {Glauber}},\ }\href@noop {} {\bibfield  {journal} {\bibinfo  {journal} {Phys.
  Rev.}\ }\textbf {\bibinfo {volume} {131}},\ \bibinfo {pages} {2766} (\bibinfo
  {year} {1963})}\BibitemShut {NoStop}%
\bibitem [{\citenamefont {Aspelmeyer}\ \emph {et~al.}(2014)\citenamefont
  {Aspelmeyer}, \citenamefont {Kippenberg},\ and\ \citenamefont
  {Marquardt}}]{aspelmeyer2014cavity}%
  \BibitemOpen
  \bibfield  {author} {\bibinfo {author} {\bibfnamefont {M.}~\bibnamefont
  {Aspelmeyer}}, \bibinfo {author} {\bibfnamefont {T.~J.}\ \bibnamefont
  {Kippenberg}}, \ and\ \bibinfo {author} {\bibfnamefont {F.}~\bibnamefont
  {Marquardt}},\ }\href@noop {} {\bibfield  {journal} {\bibinfo  {journal}
  {Rev. Mod. Phys.}\ }\textbf {\bibinfo {volume} {86}},\ \bibinfo {pages}
  {1391} (\bibinfo {year} {2014})}\BibitemShut {NoStop}%
\bibitem [{\citenamefont {Jacobs}\ \emph {et~al.}(2009)\citenamefont {Jacobs},
  \citenamefont {Tian},\ and\ \citenamefont {Finn}}]{jacobs2009engineering}%
  \BibitemOpen
  \bibfield  {author} {\bibinfo {author} {\bibfnamefont {K.}~\bibnamefont
  {Jacobs}}, \bibinfo {author} {\bibfnamefont {L.}~\bibnamefont {Tian}}, \ and\
  \bibinfo {author} {\bibfnamefont {J.}~\bibnamefont {Finn}},\ }\href@noop {}
  {\bibfield  {journal} {\bibinfo  {journal} {Phys. Rev. Lett.}\ }\textbf
  {\bibinfo {volume} {102}},\ \bibinfo {pages} {057208} (\bibinfo {year}
  {2009})}\BibitemShut {NoStop}%
\bibitem [{\citenamefont {Vanner}(2011)}]{vanner2011selective}%
  \BibitemOpen
  \bibfield  {author} {\bibinfo {author} {\bibfnamefont {M.~R.}\ \bibnamefont
  {Vanner}},\ }\href@noop {} {\bibfield  {journal} {\bibinfo  {journal} {Phys.
  Rev. X}\ }\textbf {\bibinfo {volume} {1}},\ \bibinfo {pages} {021011}
  (\bibinfo {year} {2011})}\BibitemShut {NoStop}%
\bibitem [{\citenamefont {Gr{\o}nbech-Jensen}\ \emph
  {et~al.}(2010)\citenamefont {Gr{\o}nbech-Jensen}, \citenamefont {Marchese},
  \citenamefont {Cirillo},\ and\ \citenamefont
  {Blackburn}}]{gronbech2010tomography}%
  \BibitemOpen
  \bibfield  {author} {\bibinfo {author} {\bibfnamefont {N.}~\bibnamefont
  {Gr{\o}nbech-Jensen}}, \bibinfo {author} {\bibfnamefont {J.~E.}\ \bibnamefont
  {Marchese}}, \bibinfo {author} {\bibfnamefont {M.}~\bibnamefont {Cirillo}}, \
  and\ \bibinfo {author} {\bibfnamefont {J.~A.}\ \bibnamefont {Blackburn}},\
  }\href@noop {} {\bibfield  {journal} {\bibinfo  {journal} {Phys. Rev. Lett.}\
  }\textbf {\bibinfo {volume} {105}},\ \bibinfo {pages} {010501} (\bibinfo
  {year} {2010})}\BibitemShut {NoStop}%
\bibitem [{\citenamefont {Ruggenthaler}\ and\ \citenamefont
  {Bauer}(2009)}]{ruggenthaler2009rabi}%
  \BibitemOpen
  \bibfield  {author} {\bibinfo {author} {\bibfnamefont {M.}~\bibnamefont
  {Ruggenthaler}}\ and\ \bibinfo {author} {\bibfnamefont {D.}~\bibnamefont
  {Bauer}},\ }\href@noop {} {\bibfield  {journal} {\bibinfo  {journal} {Phys.
  Rev. Lett.}\ }\textbf {\bibinfo {volume} {102}},\ \bibinfo {pages} {233001}
  (\bibinfo {year} {2009})}\BibitemShut {NoStop}%
\bibitem [{\citenamefont {Haldane}(1987)}]{haldane1987comment}%
  \BibitemOpen
  \bibfield  {author} {\bibinfo {author} {\bibfnamefont {F.}~\bibnamefont
  {Haldane}},\ }\href@noop {} {\bibfield  {journal} {\bibinfo  {journal} {Phys.
  Rev. Lett.}\ }\textbf {\bibinfo {volume} {59}},\ \bibinfo {pages} {1788}
  (\bibinfo {year} {1987})}\BibitemShut {NoStop}%
\bibitem [{\citenamefont {Segert}(1987)}]{segert1987photon}%
  \BibitemOpen
  \bibfield  {author} {\bibinfo {author} {\bibfnamefont {J.}~\bibnamefont
  {Segert}},\ }\href@noop {} {\bibfield  {journal} {\bibinfo  {journal} {Phys.
  Rev. A}\ }\textbf {\bibinfo {volume} {36}},\ \bibinfo {pages} {10} (\bibinfo
  {year} {1987})}\BibitemShut {NoStop}%
\bibitem [{\citenamefont {Mari}\ \emph {et~al.}(2011)\citenamefont {Mari},
  \citenamefont {Kieling}, \citenamefont {Nielsen}, \citenamefont {Polzik},\
  and\ \citenamefont {Eisert}}]{mari2011directly}%
  \BibitemOpen
  \bibfield  {author} {\bibinfo {author} {\bibfnamefont {A.}~\bibnamefont
  {Mari}}, \bibinfo {author} {\bibfnamefont {K.}~\bibnamefont {Kieling}},
  \bibinfo {author} {\bibfnamefont {B.~M.}\ \bibnamefont {Nielsen}}, \bibinfo
  {author} {\bibfnamefont {E.}~\bibnamefont {Polzik}}, \ and\ \bibinfo {author}
  {\bibfnamefont {J.}~\bibnamefont {Eisert}},\ }\href@noop {} {\bibfield
  {journal} {\bibinfo  {journal} {Phys. Rev. Lett.}\ }\textbf {\bibinfo
  {volume} {106}},\ \bibinfo {pages} {010403} (\bibinfo {year}
  {2011})}\BibitemShut {NoStop}%
\bibitem [{\citenamefont {Alicki}\ and\ \citenamefont
  {Van~Ryn}(2008)}]{alicki2008simple}%
  \BibitemOpen
  \bibfield  {author} {\bibinfo {author} {\bibfnamefont {R.}~\bibnamefont
  {Alicki}}\ and\ \bibinfo {author} {\bibfnamefont {N.}~\bibnamefont
  {Van~Ryn}},\ }\href@noop {} {\bibfield  {journal} {\bibinfo  {journal} {J.
  Phys. A : Math. Theor.}\ }\textbf {\bibinfo {volume} {41}},\ \bibinfo {pages}
  {062001} (\bibinfo {year} {2008})}\BibitemShut {NoStop}%
\bibitem [{\citenamefont {Richter}\ and\ \citenamefont
  {Vogel}(2002)}]{richter2002nonclassicality}%
  \BibitemOpen
  \bibfield  {author} {\bibinfo {author} {\bibfnamefont {T.}~\bibnamefont
  {Richter}}\ and\ \bibinfo {author} {\bibfnamefont {W.}~\bibnamefont
  {Vogel}},\ }\href@noop {} {\bibfield  {journal} {\bibinfo  {journal} {Phys.
  Rev. Lett.}\ }\textbf {\bibinfo {volume} {89}},\ \bibinfo {pages} {283601}
  (\bibinfo {year} {2002})}\BibitemShut {NoStop}%
\bibitem [{\citenamefont {Pikovski}\ \emph {et~al.}(2012)\citenamefont
  {Pikovski}, \citenamefont {Vanner}, \citenamefont {Aspelmeyer}, \citenamefont
  {Kim},\ and\ \citenamefont {Brukner}}]{pikovski2012probing}%
  \BibitemOpen
  \bibfield  {author} {\bibinfo {author} {\bibfnamefont {I.}~\bibnamefont
  {Pikovski}}, \bibinfo {author} {\bibfnamefont {M.~R.}\ \bibnamefont
  {Vanner}}, \bibinfo {author} {\bibfnamefont {M.}~\bibnamefont {Aspelmeyer}},
  \bibinfo {author} {\bibfnamefont {M.~S.}\ \bibnamefont {Kim}}, \ and\
  \bibinfo {author} {\bibfnamefont {{\v{C}}.}~\bibnamefont {Brukner}},\
  }\href@noop {} {\bibfield  {journal} {\bibinfo  {journal} {Nat. Phys.}\
  }\textbf {\bibinfo {volume} {8}},\ \bibinfo {pages} {393} (\bibinfo {year}
  {2012})}\BibitemShut {NoStop}%
\bibitem [{\citenamefont {Penrose}(1998)}]{penrose1998quantum}%
  \BibitemOpen
  \bibfield  {author} {\bibinfo {author} {\bibfnamefont {R.}~\bibnamefont
  {Penrose}},\ }\href@noop {} {\bibfield  {journal} {\bibinfo  {journal} {Phil.
  Trans. R. Soc. A}\ }\textbf {\bibinfo {volume} {356}},\ \bibinfo {pages}
  {1927} (\bibinfo {year} {1998})}\BibitemShut {NoStop}%
\bibitem [{\citenamefont {Marshall}\ \emph {et~al.}(2003)\citenamefont
  {Marshall}, \citenamefont {Simon}, \citenamefont {Penrose},\ and\
  \citenamefont {Bouwmeester}}]{marshall2003towards}%
  \BibitemOpen
  \bibfield  {author} {\bibinfo {author} {\bibfnamefont {W.}~\bibnamefont
  {Marshall}}, \bibinfo {author} {\bibfnamefont {C.}~\bibnamefont {Simon}},
  \bibinfo {author} {\bibfnamefont {R.}~\bibnamefont {Penrose}}, \ and\
  \bibinfo {author} {\bibfnamefont {D.}~\bibnamefont {Bouwmeester}},\
  }\href@noop {} {\bibfield  {journal} {\bibinfo  {journal} {Phys. Rev. Lett.}\
  }\textbf {\bibinfo {volume} {91}},\ \bibinfo {pages} {130401} (\bibinfo
  {year} {2003})}\BibitemShut {NoStop}%
\bibitem [{\citenamefont {Penrose}(1996)}]{penrose1996gravity}%
  \BibitemOpen
  \bibfield  {author} {\bibinfo {author} {\bibfnamefont {R.}~\bibnamefont
  {Penrose}},\ }\href@noop {} {\bibfield  {journal} {\bibinfo  {journal} {Gen.
  Rel. Gravit.}\ }\textbf {\bibinfo {volume} {28}},\ \bibinfo {pages} {581}
  (\bibinfo {year} {1996})}\BibitemShut {NoStop}%
\bibitem [{\citenamefont {Di{\'o}si}(1989)}]{diosi1989models}%
  \BibitemOpen
  \bibfield  {author} {\bibinfo {author} {\bibfnamefont {L.}~\bibnamefont
  {Di{\'o}si}},\ }\href@noop {} {\bibfield  {journal} {\bibinfo  {journal}
  {Phys. Rev. A}\ }\textbf {\bibinfo {volume} {40}},\ \bibinfo {pages} {1165}
  (\bibinfo {year} {1989})}\BibitemShut {NoStop}%
\bibitem [{\citenamefont {Schm{\"o}le}\ \emph {et~al.}(2016)\citenamefont
  {Schm{\"o}le}, \citenamefont {Dragosits}, \citenamefont {Hepach},\ and\
  \citenamefont {Aspelmeyer}}]{schmole2016micromechanical}%
  \BibitemOpen
  \bibfield  {author} {\bibinfo {author} {\bibfnamefont {J.}~\bibnamefont
  {Schm{\"o}le}}, \bibinfo {author} {\bibfnamefont {M.}~\bibnamefont
  {Dragosits}}, \bibinfo {author} {\bibfnamefont {H.}~\bibnamefont {Hepach}}, \
  and\ \bibinfo {author} {\bibfnamefont {M.}~\bibnamefont {Aspelmeyer}},\
  }\href@noop {} {\bibfield  {journal} {\bibinfo  {journal} {Class. Quantum
  Grav.}\ }\textbf {\bibinfo {volume} {33}},\ \bibinfo {pages} {125031}
  (\bibinfo {year} {2016})}\BibitemShut {NoStop}%
\bibitem [{\citenamefont {Plato}\ \emph {et~al.}(2016)\citenamefont {Plato},
  \citenamefont {Hughes},\ and\ \citenamefont {Kim}}]{plato2016gravitational}%
  \BibitemOpen
  \bibfield  {author} {\bibinfo {author} {\bibfnamefont {A.}~\bibnamefont
  {Plato}}, \bibinfo {author} {\bibfnamefont {C.}~\bibnamefont {Hughes}}, \
  and\ \bibinfo {author} {\bibfnamefont {M.~S.}\ \bibnamefont {Kim}},\
  }\href@noop {} {\bibfield  {journal} {\bibinfo  {journal} {Contemp. Phys.}\
  }\textbf {\bibinfo {volume} {57}},\ \bibinfo {pages} {477} (\bibinfo {year}
  {2016})}\BibitemShut {NoStop}%
\bibitem [{\citenamefont {Latmiral}\ \emph {et~al.}(2016)\citenamefont
  {Latmiral}, \citenamefont {Armata}, \citenamefont {Genoni}, \citenamefont
  {Pikovski},\ and\ \citenamefont {Kim}}]{latmiral2016probing}%
  \BibitemOpen
  \bibfield  {author} {\bibinfo {author} {\bibfnamefont {L.}~\bibnamefont
  {Latmiral}}, \bibinfo {author} {\bibfnamefont {F.}~\bibnamefont {Armata}},
  \bibinfo {author} {\bibfnamefont {M.~G.}\ \bibnamefont {Genoni}}, \bibinfo
  {author} {\bibfnamefont {I.}~\bibnamefont {Pikovski}}, \ and\ \bibinfo
  {author} {\bibfnamefont {M.~S.}\ \bibnamefont {Kim}},\ }\href@noop {}
  {\bibfield  {journal} {\bibinfo  {journal} {Phys. Rev. A}\ }\textbf {\bibinfo
  {volume} {93}},\ \bibinfo {pages} {052306} (\bibinfo {year}
  {2016})}\BibitemShut {NoStop}%
\bibitem [{\citenamefont {Gavartin}\ \emph {et~al.}(2012)\citenamefont
  {Gavartin}, \citenamefont {Verlot},\ and\ \citenamefont
  {Kippenberg}}]{gavartin2012hybrid}%
  \BibitemOpen
  \bibfield  {author} {\bibinfo {author} {\bibfnamefont {E.}~\bibnamefont
  {Gavartin}}, \bibinfo {author} {\bibfnamefont {P.}~\bibnamefont {Verlot}}, \
  and\ \bibinfo {author} {\bibfnamefont {T.~J.}\ \bibnamefont {Kippenberg}},\
  }\href@noop {} {\bibfield  {journal} {\bibinfo  {journal} {Nat.
  Nanotechnol.}\ }\textbf {\bibinfo {volume} {7}},\ \bibinfo {pages} {509}
  (\bibinfo {year} {2012})}\BibitemShut {NoStop}%
\bibitem [{\citenamefont {Forstner}\ \emph {et~al.}(2012)\citenamefont
  {Forstner}, \citenamefont {Prams}, \citenamefont {Knittel}, \citenamefont
  {Van~Ooijen}, \citenamefont {Swaim}, \citenamefont {Harris}, \citenamefont
  {Szorkovszky}, \citenamefont {Bowen},\ and\ \citenamefont
  {Rubinsztein-Dunlop}}]{forstner2012cavity}%
  \BibitemOpen
  \bibfield  {author} {\bibinfo {author} {\bibfnamefont {S.}~\bibnamefont
  {Forstner}}, \bibinfo {author} {\bibfnamefont {S.}~\bibnamefont {Prams}},
  \bibinfo {author} {\bibfnamefont {J.}~\bibnamefont {Knittel}}, \bibinfo
  {author} {\bibfnamefont {E.}~\bibnamefont {Van~Ooijen}}, \bibinfo {author}
  {\bibfnamefont {J.}~\bibnamefont {Swaim}}, \bibinfo {author} {\bibfnamefont
  {G.}~\bibnamefont {Harris}}, \bibinfo {author} {\bibfnamefont
  {A.}~\bibnamefont {Szorkovszky}}, \bibinfo {author} {\bibfnamefont
  {W.}~\bibnamefont {Bowen}}, \ and\ \bibinfo {author} {\bibfnamefont
  {H.}~\bibnamefont {Rubinsztein-Dunlop}},\ }\href@noop {} {\bibfield
  {journal} {\bibinfo  {journal} {Phys. Rev. Lett.}\ }\textbf {\bibinfo
  {volume} {108}},\ \bibinfo {pages} {120801} (\bibinfo {year}
  {2012})}\BibitemShut {NoStop}%
\bibitem [{\citenamefont {Krause}\ \emph {et~al.}(2012)\citenamefont {Krause},
  \citenamefont {Winger}, \citenamefont {Blasius}, \citenamefont {Lin},\ and\
  \citenamefont {Painter}}]{krause2012high}%
  \BibitemOpen
  \bibfield  {author} {\bibinfo {author} {\bibfnamefont {A.~G.}\ \bibnamefont
  {Krause}}, \bibinfo {author} {\bibfnamefont {M.}~\bibnamefont {Winger}},
  \bibinfo {author} {\bibfnamefont {T.~D.}\ \bibnamefont {Blasius}}, \bibinfo
  {author} {\bibfnamefont {Q.}~\bibnamefont {Lin}}, \ and\ \bibinfo {author}
  {\bibfnamefont {O.}~\bibnamefont {Painter}},\ }\href@noop {} {\bibfield
  {journal} {\bibinfo  {journal} {Nat. Photonics}\ }\textbf {\bibinfo {volume}
  {6}},\ \bibinfo {pages} {768} (\bibinfo {year} {2012})}\BibitemShut {NoStop}%
\bibitem [{\citenamefont {Bagci}\ \emph {et~al.}(2014)\citenamefont {Bagci},
  \citenamefont {Simonsen}, \citenamefont {Schmid}, \citenamefont {Villanueva},
  \citenamefont {Zeuthen}, \citenamefont {Appel}, \citenamefont {Taylor},
  \citenamefont {S{\o}rensen}, \citenamefont {Usami}, \citenamefont
  {Schliesser} \emph {et~al.}}]{bagci2014optical}%
  \BibitemOpen
  \bibfield  {author} {\bibinfo {author} {\bibfnamefont {T.}~\bibnamefont
  {Bagci}}, \bibinfo {author} {\bibfnamefont {A.}~\bibnamefont {Simonsen}},
  \bibinfo {author} {\bibfnamefont {S.}~\bibnamefont {Schmid}}, \bibinfo
  {author} {\bibfnamefont {L.~G.}\ \bibnamefont {Villanueva}}, \bibinfo
  {author} {\bibfnamefont {E.}~\bibnamefont {Zeuthen}}, \bibinfo {author}
  {\bibfnamefont {J.}~\bibnamefont {Appel}}, \bibinfo {author} {\bibfnamefont
  {J.~M.}\ \bibnamefont {Taylor}}, \bibinfo {author} {\bibfnamefont
  {A.}~\bibnamefont {S{\o}rensen}}, \bibinfo {author} {\bibfnamefont
  {K.}~\bibnamefont {Usami}}, \bibinfo {author} {\bibfnamefont
  {A.}~\bibnamefont {Schliesser}},  \emph {et~al.},\ }\href@noop {} {\bibfield
  {journal} {\bibinfo  {journal} {Nature}\ }\textbf {\bibinfo {volume} {507}},\
  \bibinfo {pages} {81} (\bibinfo {year} {2014})}\BibitemShut {NoStop}%
\bibitem [{\citenamefont {Vanner}\ \emph {et~al.}(2011)\citenamefont {Vanner},
  \citenamefont {Pikovski}, \citenamefont {Cole}, \citenamefont {Kim},
  \citenamefont {Brukner}, \citenamefont {Hammerer}, \citenamefont {Milburn},\
  and\ \citenamefont {Aspelmeyer}}]{vanner2011pulsed}%
  \BibitemOpen
  \bibfield  {author} {\bibinfo {author} {\bibfnamefont {M.~R.}\ \bibnamefont
  {Vanner}}, \bibinfo {author} {\bibfnamefont {I.}~\bibnamefont {Pikovski}},
  \bibinfo {author} {\bibfnamefont {G.~D.}\ \bibnamefont {Cole}}, \bibinfo
  {author} {\bibfnamefont {M.~S.}\ \bibnamefont {Kim}}, \bibinfo {author}
  {\bibfnamefont {{\v{C}}.}~\bibnamefont {Brukner}}, \bibinfo {author}
  {\bibfnamefont {K.}~\bibnamefont {Hammerer}}, \bibinfo {author}
  {\bibfnamefont {G.~J.}\ \bibnamefont {Milburn}}, \ and\ \bibinfo {author}
  {\bibfnamefont {M.}~\bibnamefont {Aspelmeyer}},\ }\href@noop {} {\bibfield
  {journal} {\bibinfo  {journal} {Proc. Natl. Acad. Sci.}\ }\textbf {\bibinfo
  {volume} {108}},\ \bibinfo {pages} {16182} (\bibinfo {year}
  {2011})}\BibitemShut {NoStop}%
\bibitem [{\citenamefont {Vanner}\ \emph {et~al.}(2015)\citenamefont {Vanner},
  \citenamefont {Pikovski},\ and\ \citenamefont {Kim}}]{vanner2015towards}%
  \BibitemOpen
  \bibfield  {author} {\bibinfo {author} {\bibfnamefont {M.~R.}\ \bibnamefont
  {Vanner}}, \bibinfo {author} {\bibfnamefont {I.}~\bibnamefont {Pikovski}}, \
  and\ \bibinfo {author} {\bibfnamefont {M.~S.}\ \bibnamefont {Kim}},\
  }\href@noop {} {\bibfield  {journal} {\bibinfo  {journal} {Ann. Phys.
  (Berl.)}\ }\textbf {\bibinfo {volume} {527}},\ \bibinfo {pages} {15}
  (\bibinfo {year} {2015})}\BibitemShut {NoStop}%
\bibitem [{\citenamefont {Safavi-Naeini}\ \emph {et~al.}(2012)\citenamefont
  {Safavi-Naeini}, \citenamefont {Chan}, \citenamefont {Hill}, \citenamefont
  {Alegre}, \citenamefont {Krause},\ and\ \citenamefont
  {Painter}}]{safavi2012observation}%
  \BibitemOpen
  \bibfield  {author} {\bibinfo {author} {\bibfnamefont {A.~H.}\ \bibnamefont
  {Safavi-Naeini}}, \bibinfo {author} {\bibfnamefont {J.}~\bibnamefont {Chan}},
  \bibinfo {author} {\bibfnamefont {J.~T.}\ \bibnamefont {Hill}}, \bibinfo
  {author} {\bibfnamefont {T.~P.~M.}\ \bibnamefont {Alegre}}, \bibinfo {author}
  {\bibfnamefont {A.}~\bibnamefont {Krause}}, \ and\ \bibinfo {author}
  {\bibfnamefont {O.}~\bibnamefont {Painter}},\ }\href@noop {} {\bibfield
  {journal} {\bibinfo  {journal} {Phys. Rev. Lett.}\ }\textbf {\bibinfo
  {volume} {108}},\ \bibinfo {pages} {033602} (\bibinfo {year}
  {2012})}\BibitemShut {NoStop}%
\bibitem [{\citenamefont {Weinstein}\ \emph {et~al.}(2014)\citenamefont
  {Weinstein}, \citenamefont {Lei}, \citenamefont {Wollman}, \citenamefont
  {Suh}, \citenamefont {Metelmann}, \citenamefont {Clerk},\ and\ \citenamefont
  {Schwab}}]{weinstein2014observation}%
  \BibitemOpen
  \bibfield  {author} {\bibinfo {author} {\bibfnamefont {A.}~\bibnamefont
  {Weinstein}}, \bibinfo {author} {\bibfnamefont {C.}~\bibnamefont {Lei}},
  \bibinfo {author} {\bibfnamefont {E.}~\bibnamefont {Wollman}}, \bibinfo
  {author} {\bibfnamefont {J.}~\bibnamefont {Suh}}, \bibinfo {author}
  {\bibfnamefont {A.}~\bibnamefont {Metelmann}}, \bibinfo {author}
  {\bibfnamefont {A.}~\bibnamefont {Clerk}}, \ and\ \bibinfo {author}
  {\bibfnamefont {K.}~\bibnamefont {Schwab}},\ }\href@noop {} {\bibfield
  {journal} {\bibinfo  {journal} {Phys. Rev. X}\ }\textbf {\bibinfo {volume}
  {4}},\ \bibinfo {pages} {041003} (\bibinfo {year} {2014})}\BibitemShut
  {NoStop}%
\bibitem [{\citenamefont {Sudhir}\ \emph {et~al.}(2017)\citenamefont {Sudhir},
  \citenamefont {Wilson}, \citenamefont {Schilling}, \citenamefont
  {Sch{\"u}tz}, \citenamefont {Fedorov}, \citenamefont {Ghadimi}, \citenamefont
  {Nunnenkamp},\ and\ \citenamefont {Kippenberg}}]{sudhir2017appearance}%
  \BibitemOpen
  \bibfield  {author} {\bibinfo {author} {\bibfnamefont {V.}~\bibnamefont
  {Sudhir}}, \bibinfo {author} {\bibfnamefont {D.~J.}\ \bibnamefont {Wilson}},
  \bibinfo {author} {\bibfnamefont {R.}~\bibnamefont {Schilling}}, \bibinfo
  {author} {\bibfnamefont {H.}~\bibnamefont {Sch{\"u}tz}}, \bibinfo {author}
  {\bibfnamefont {S.~A.}\ \bibnamefont {Fedorov}}, \bibinfo {author}
  {\bibfnamefont {A.~H.}\ \bibnamefont {Ghadimi}}, \bibinfo {author}
  {\bibfnamefont {A.}~\bibnamefont {Nunnenkamp}}, \ and\ \bibinfo {author}
  {\bibfnamefont {T.~J.}\ \bibnamefont {Kippenberg}},\ }\href@noop {}
  {\bibfield  {journal} {\bibinfo  {journal} {Phys. Rev. X}\ }\textbf {\bibinfo
  {volume} {7}},\ \bibinfo {pages} {011001} (\bibinfo {year}
  {2017})}\BibitemShut {NoStop}%
\bibitem [{\citenamefont {Armata}\ \emph {et~al.}(2016)\citenamefont {Armata},
  \citenamefont {Latmiral}, \citenamefont {Pikovski}, \citenamefont {Vanner},
  \citenamefont {Brukner},\ and\ \citenamefont {Kim}}]{armata2016quantum}%
  \BibitemOpen
  \bibfield  {author} {\bibinfo {author} {\bibfnamefont {F.}~\bibnamefont
  {Armata}}, \bibinfo {author} {\bibfnamefont {L.}~\bibnamefont {Latmiral}},
  \bibinfo {author} {\bibfnamefont {I.}~\bibnamefont {Pikovski}}, \bibinfo
  {author} {\bibfnamefont {M.~R.}\ \bibnamefont {Vanner}}, \bibinfo {author}
  {\bibfnamefont {{\v{C}}.}~\bibnamefont {Brukner}}, \ and\ \bibinfo {author}
  {\bibfnamefont {M.~S.}\ \bibnamefont {Kim}},\ }\href@noop {} {\bibfield
  {journal} {\bibinfo  {journal} {Phys. Rev. A}\ }\textbf {\bibinfo {volume}
  {93}},\ \bibinfo {pages} {063862} (\bibinfo {year} {2016})}\BibitemShut
  {NoStop}%
\bibitem [{\citenamefont {Fabre}\ \emph {et~al.}(1994)\citenamefont {Fabre},
  \citenamefont {Pinard}, \citenamefont {Bourzeix}, \citenamefont {Heidmann},
  \citenamefont {Giacobino},\ and\ \citenamefont {Reynaud}}]{fabre1994quantum}%
  \BibitemOpen
  \bibfield  {author} {\bibinfo {author} {\bibfnamefont {C.}~\bibnamefont
  {Fabre}}, \bibinfo {author} {\bibfnamefont {M.}~\bibnamefont {Pinard}},
  \bibinfo {author} {\bibfnamefont {S.}~\bibnamefont {Bourzeix}}, \bibinfo
  {author} {\bibfnamefont {A.}~\bibnamefont {Heidmann}}, \bibinfo {author}
  {\bibfnamefont {E.}~\bibnamefont {Giacobino}}, \ and\ \bibinfo {author}
  {\bibfnamefont {S.}~\bibnamefont {Reynaud}},\ }\href@noop {} {\bibfield
  {journal} {\bibinfo  {journal} {Phys. Rev. A}\ }\textbf {\bibinfo {volume}
  {49}},\ \bibinfo {pages} {1337} (\bibinfo {year} {1994})}\BibitemShut
  {NoStop}%
\bibitem [{\citenamefont {Mancini}\ and\ \citenamefont
  {Tombesi}(1994)}]{mancini1994quantum}%
  \BibitemOpen
  \bibfield  {author} {\bibinfo {author} {\bibfnamefont {S.}~\bibnamefont
  {Mancini}}\ and\ \bibinfo {author} {\bibfnamefont {P.}~\bibnamefont
  {Tombesi}},\ }\href@noop {} {\bibfield  {journal} {\bibinfo  {journal} {Phys.
  Rev. A}\ }\textbf {\bibinfo {volume} {49}},\ \bibinfo {pages} {4055}
  (\bibinfo {year} {1994})}\BibitemShut {NoStop}%
\bibitem [{\citenamefont {Brooks}\ \emph {et~al.}(2012)\citenamefont {Brooks},
  \citenamefont {Botter}, \citenamefont {Schreppler}, \citenamefont {Purdy},
  \citenamefont {Brahms},\ and\ \citenamefont {Stamper-Kurn}}]{brooks2012non}%
  \BibitemOpen
  \bibfield  {author} {\bibinfo {author} {\bibfnamefont {D.~W.}\ \bibnamefont
  {Brooks}}, \bibinfo {author} {\bibfnamefont {T.}~\bibnamefont {Botter}},
  \bibinfo {author} {\bibfnamefont {S.}~\bibnamefont {Schreppler}}, \bibinfo
  {author} {\bibfnamefont {T.~P.}\ \bibnamefont {Purdy}}, \bibinfo {author}
  {\bibfnamefont {N.}~\bibnamefont {Brahms}}, \ and\ \bibinfo {author}
  {\bibfnamefont {D.~M.}\ \bibnamefont {Stamper-Kurn}},\ }\href@noop {}
  {\bibfield  {journal} {\bibinfo  {journal} {Nature}\ }\textbf {\bibinfo
  {volume} {488}},\ \bibinfo {pages} {476} (\bibinfo {year}
  {2012})}\BibitemShut {NoStop}%
\bibitem [{\citenamefont {Purdy}\ \emph {et~al.}(2013)\citenamefont {Purdy},
  \citenamefont {Yu}, \citenamefont {Peterson}, \citenamefont {Kampel},\ and\
  \citenamefont {Regal}}]{purdy2013strong}%
  \BibitemOpen
  \bibfield  {author} {\bibinfo {author} {\bibfnamefont {T.}~\bibnamefont
  {Purdy}}, \bibinfo {author} {\bibfnamefont {P.-L.}\ \bibnamefont {Yu}},
  \bibinfo {author} {\bibfnamefont {R.}~\bibnamefont {Peterson}}, \bibinfo
  {author} {\bibfnamefont {N.}~\bibnamefont {Kampel}}, \ and\ \bibinfo {author}
  {\bibfnamefont {C.}~\bibnamefont {Regal}},\ }\href@noop {} {\bibfield
  {journal} {\bibinfo  {journal} {Phys. Rev. X}\ }\textbf {\bibinfo {volume}
  {3}},\ \bibinfo {pages} {031012} (\bibinfo {year} {2013})}\BibitemShut
  {NoStop}%
\bibitem [{\citenamefont {Safavi-Naeini}\ \emph {et~al.}(2013)\citenamefont
  {Safavi-Naeini}, \citenamefont {Gr{\"o}blacher}, \citenamefont {Hill},
  \citenamefont {Chan}, \citenamefont {Aspelmeyer},\ and\ \citenamefont
  {Painter}}]{safavi2013squeezed}%
  \BibitemOpen
  \bibfield  {author} {\bibinfo {author} {\bibfnamefont {A.~H.}\ \bibnamefont
  {Safavi-Naeini}}, \bibinfo {author} {\bibfnamefont {S.}~\bibnamefont
  {Gr{\"o}blacher}}, \bibinfo {author} {\bibfnamefont {J.~T.}\ \bibnamefont
  {Hill}}, \bibinfo {author} {\bibfnamefont {J.}~\bibnamefont {Chan}}, \bibinfo
  {author} {\bibfnamefont {M.}~\bibnamefont {Aspelmeyer}}, \ and\ \bibinfo
  {author} {\bibfnamefont {O.}~\bibnamefont {Painter}},\ }\href@noop {}
  {\bibfield  {journal} {\bibinfo  {journal} {Nature}\ }\textbf {\bibinfo
  {volume} {500}},\ \bibinfo {pages} {185} (\bibinfo {year}
  {2013})}\BibitemShut {NoStop}%
\bibitem [{\citenamefont {Dellantonio}\ \emph {et~al.}(2018)\citenamefont
  {Dellantonio}, \citenamefont {Kyriienko}, \citenamefont {Marquardt},\ and\
  \citenamefont {S{\o}rensen}}]{dellantonio2018quantum}%
  \BibitemOpen
  \bibfield  {author} {\bibinfo {author} {\bibfnamefont {L.}~\bibnamefont
  {Dellantonio}}, \bibinfo {author} {\bibfnamefont {O.}~\bibnamefont
  {Kyriienko}}, \bibinfo {author} {\bibfnamefont {F.}~\bibnamefont
  {Marquardt}}, \ and\ \bibinfo {author} {\bibfnamefont {A.~S.}\ \bibnamefont
  {S{\o}rensen}},\ }\href@noop {} {\bibfield  {journal} {\bibinfo  {journal}
  {Nat. Commun.}\ }\textbf {\bibinfo {volume} {9}},\ \bibinfo {pages} {3621}
  (\bibinfo {year} {2018})}\BibitemShut {NoStop}%
\bibitem [{\citenamefont {Lloyd}\ and\ \citenamefont
  {Braunstein}(1999)}]{lloyd1999quantum}%
  \BibitemOpen
  \bibfield  {author} {\bibinfo {author} {\bibfnamefont {S.}~\bibnamefont
  {Lloyd}}\ and\ \bibinfo {author} {\bibfnamefont {S.~L.}\ \bibnamefont
  {Braunstein}},\ }in\ \href@noop {} {\emph {\bibinfo {booktitle} {Quantum
  Information with Continuous Variables}}}\ (\bibinfo  {publisher} {Springer},\
  \bibinfo {year} {1999})\ pp.\ \bibinfo {pages} {9--17}\BibitemShut {NoStop}%
\bibitem [{\citenamefont {Pace}\ \emph {et~al.}(1993)\citenamefont {Pace},
  \citenamefont {Collett},\ and\ \citenamefont {Walls}}]{pace1993quantum}%
  \BibitemOpen
  \bibfield  {author} {\bibinfo {author} {\bibfnamefont {A.}~\bibnamefont
  {Pace}}, \bibinfo {author} {\bibfnamefont {M.}~\bibnamefont {Collett}}, \
  and\ \bibinfo {author} {\bibfnamefont {D.}~\bibnamefont {Walls}},\
  }\href@noop {} {\bibfield  {journal} {\bibinfo  {journal} {Phys. Rev. A}\
  }\textbf {\bibinfo {volume} {47}},\ \bibinfo {pages} {3173} (\bibinfo {year}
  {1993})}\BibitemShut {NoStop}%
\bibitem [{\citenamefont {Law}(1995)}]{law1995interaction}%
  \BibitemOpen
  \bibfield  {author} {\bibinfo {author} {\bibfnamefont {C.}~\bibnamefont
  {Law}},\ }\href@noop {} {\bibfield  {journal} {\bibinfo  {journal} {Phys.
  Rev. A}\ }\textbf {\bibinfo {volume} {51}},\ \bibinfo {pages} {2537}
  (\bibinfo {year} {1995})}\BibitemShut {NoStop}%
\bibitem [{\citenamefont {Mancini}\ \emph {et~al.}(1997)\citenamefont
  {Mancini}, \citenamefont {Man'ko},\ and\ \citenamefont
  {Tombesi}}]{mancini1997ponderomotive}%
  \BibitemOpen
  \bibfield  {author} {\bibinfo {author} {\bibfnamefont {S.}~\bibnamefont
  {Mancini}}, \bibinfo {author} {\bibfnamefont {V.}~\bibnamefont {Man'ko}}, \
  and\ \bibinfo {author} {\bibfnamefont {P.}~\bibnamefont {Tombesi}},\
  }\href@noop {} {\bibfield  {journal} {\bibinfo  {journal} {Phys. Rev. A}\
  }\textbf {\bibinfo {volume} {55}},\ \bibinfo {pages} {3042} (\bibinfo {year}
  {1997})}\BibitemShut {NoStop}%
\bibitem [{sup()}]{supp}%
  \BibitemOpen
  \href@noop {} {}\bibinfo {howpublished} {See supplementary
  material}\BibitemShut {NoStop}%
\bibitem [{\citenamefont {Baker}\ \emph {et~al.}(2016)\citenamefont {Baker},
  \citenamefont {Harris}, \citenamefont {McAuslan}, \citenamefont {Sachkou},
  \citenamefont {He},\ and\ \citenamefont {Bowen}}]{baker2016theoretical}%
  \BibitemOpen
  \bibfield  {author} {\bibinfo {author} {\bibfnamefont {C.~G.}\ \bibnamefont
  {Baker}}, \bibinfo {author} {\bibfnamefont {G.~I.}\ \bibnamefont {Harris}},
  \bibinfo {author} {\bibfnamefont {D.~L.}\ \bibnamefont {McAuslan}}, \bibinfo
  {author} {\bibfnamefont {Y.}~\bibnamefont {Sachkou}}, \bibinfo {author}
  {\bibfnamefont {X.}~\bibnamefont {He}}, \ and\ \bibinfo {author}
  {\bibfnamefont {W.~P.}\ \bibnamefont {Bowen}},\ }\href@noop {} {\bibfield
  {journal} {\bibinfo  {journal} {New J. Phys.}\ }\textbf {\bibinfo {volume}
  {18}},\ \bibinfo {pages} {123025} (\bibinfo {year} {2016})}\BibitemShut
  {NoStop}%
\bibitem [{\citenamefont {Milburn}(1986)}]{milburn1986quantum}%
  \BibitemOpen
  \bibfield  {author} {\bibinfo {author} {\bibfnamefont {G.}~\bibnamefont
  {Milburn}},\ }\href@noop {} {\bibfield  {journal} {\bibinfo  {journal} {Phys.
  Rev. A}\ }\textbf {\bibinfo {volume} {33}},\ \bibinfo {pages} {674} (\bibinfo
  {year} {1986})}\BibitemShut {NoStop}%
\bibitem [{\citenamefont {James}\ and\ \citenamefont
  {Jerke}(2007)}]{james2007effective}%
  \BibitemOpen
  \bibfield  {author} {\bibinfo {author} {\bibfnamefont {D.}~\bibnamefont
  {James}}\ and\ \bibinfo {author} {\bibfnamefont {J.}~\bibnamefont {Jerke}},\
  }\href@noop {} {\bibfield  {journal} {\bibinfo  {journal} {Can. J. Phys.}\
  }\textbf {\bibinfo {volume} {85}},\ \bibinfo {pages} {625} (\bibinfo {year}
  {2007})}\BibitemShut {NoStop}%
\bibitem [{\citenamefont {Marshall}\ and\ \citenamefont
  {Santos}(1988)}]{marshall1988stochastic}%
  \BibitemOpen
  \bibfield  {author} {\bibinfo {author} {\bibfnamefont {T.}~\bibnamefont
  {Marshall}}\ and\ \bibinfo {author} {\bibfnamefont {E.}~\bibnamefont
  {Santos}},\ }\href@noop {} {\bibfield  {journal} {\bibinfo  {journal} {Found.
  Phys.}\ }\textbf {\bibinfo {volume} {18}},\ \bibinfo {pages} {185} (\bibinfo
  {year} {1988})}\BibitemShut {NoStop}%
\bibitem [{\citenamefont {Murch}\ \emph {et~al.}(2008)\citenamefont {Murch},
  \citenamefont {Moore}, \citenamefont {Gupta},\ and\ \citenamefont
  {Stamper-Kurn}}]{murch2008observation}%
  \BibitemOpen
  \bibfield  {author} {\bibinfo {author} {\bibfnamefont {K.~W.}\ \bibnamefont
  {Murch}}, \bibinfo {author} {\bibfnamefont {K.~L.}\ \bibnamefont {Moore}},
  \bibinfo {author} {\bibfnamefont {S.}~\bibnamefont {Gupta}}, \ and\ \bibinfo
  {author} {\bibfnamefont {D.~M.}\ \bibnamefont {Stamper-Kurn}},\ }\href@noop
  {} {\bibfield  {journal} {\bibinfo  {journal} {Nat. Phys.}\ }\textbf
  {\bibinfo {volume} {4}},\ \bibinfo {pages} {561} (\bibinfo {year}
  {2008})}\BibitemShut {NoStop}%
\bibitem [{\citenamefont {Brennecke}\ \emph {et~al.}(2008)\citenamefont
  {Brennecke}, \citenamefont {Ritter}, \citenamefont {Donner},\ and\
  \citenamefont {Esslinger}}]{brennecke2008cavity}%
  \BibitemOpen
  \bibfield  {author} {\bibinfo {author} {\bibfnamefont {F.}~\bibnamefont
  {Brennecke}}, \bibinfo {author} {\bibfnamefont {S.}~\bibnamefont {Ritter}},
  \bibinfo {author} {\bibfnamefont {T.}~\bibnamefont {Donner}}, \ and\ \bibinfo
  {author} {\bibfnamefont {T.}~\bibnamefont {Esslinger}},\ }\href@noop {}
  {\bibfield  {journal} {\bibinfo  {journal} {Science}\ }\textbf {\bibinfo
  {volume} {322}},\ \bibinfo {pages} {235} (\bibinfo {year}
  {2008})}\BibitemShut {NoStop}%
\bibitem [{\citenamefont {Eppley}\ and\ \citenamefont
  {Hannah}(1977)}]{eppley1977necessity}%
  \BibitemOpen
  \bibfield  {author} {\bibinfo {author} {\bibfnamefont {K.}~\bibnamefont
  {Eppley}}\ and\ \bibinfo {author} {\bibfnamefont {E.}~\bibnamefont
  {Hannah}},\ }\href@noop {} {\bibfield  {journal} {\bibinfo  {journal} {Found.
  Phys.}\ }\textbf {\bibinfo {volume} {7}},\ \bibinfo {pages} {51} (\bibinfo
  {year} {1977})}\BibitemShut {NoStop}%
\bibitem [{\citenamefont {Di{\'o}si}\ \emph {et~al.}(2000)\citenamefont
  {Di{\'o}si}, \citenamefont {Gisin},\ and\ \citenamefont
  {Strunz}}]{diosi2000quantum}%
  \BibitemOpen
  \bibfield  {author} {\bibinfo {author} {\bibfnamefont {L.}~\bibnamefont
  {Di{\'o}si}}, \bibinfo {author} {\bibfnamefont {N.}~\bibnamefont {Gisin}}, \
  and\ \bibinfo {author} {\bibfnamefont {W.~T.}\ \bibnamefont {Strunz}},\
  }\href@noop {} {\bibfield  {journal} {\bibinfo  {journal} {Phys. Rev. A}\
  }\textbf {\bibinfo {volume} {61}},\ \bibinfo {pages} {022108} (\bibinfo
  {year} {2000})}\BibitemShut {NoStop}%
\bibitem [{\citenamefont {Adler}(1994)}]{adler1994generalized}%
  \BibitemOpen
  \bibfield  {author} {\bibinfo {author} {\bibfnamefont {S.~L.}\ \bibnamefont
  {Adler}},\ }\href@noop {} {\bibfield  {journal} {\bibinfo  {journal} {Nucl.
  Phys. B}\ }\textbf {\bibinfo {volume} {415}},\ \bibinfo {pages} {195}
  (\bibinfo {year} {1994})}\BibitemShut {NoStop}%
\bibitem [{\citenamefont {Elze}(2012)}]{elze2012linear}%
  \BibitemOpen
  \bibfield  {author} {\bibinfo {author} {\bibfnamefont {H.-T.}\ \bibnamefont
  {Elze}},\ }\href@noop {} {\bibfield  {journal} {\bibinfo  {journal} {Phys.
  Rev. A}\ }\textbf {\bibinfo {volume} {85}},\ \bibinfo {pages} {052109}
  (\bibinfo {year} {2012})}\BibitemShut {NoStop}%
\bibitem [{\citenamefont {Di{\'o}si}(1988)}]{diosi1988continuous}%
  \BibitemOpen
  \bibfield  {author} {\bibinfo {author} {\bibfnamefont {L.}~\bibnamefont
  {Di{\'o}si}},\ }\href@noop {} {\bibfield  {journal} {\bibinfo  {journal}
  {Phys. Lett. A}\ }\textbf {\bibinfo {volume} {129}},\ \bibinfo {pages} {419}
  (\bibinfo {year} {1988})}\BibitemShut {NoStop}%
\bibitem [{\citenamefont {Jacobs}\ and\ \citenamefont
  {Steck}(2006)}]{jacobs2006straightforward}%
  \BibitemOpen
  \bibfield  {author} {\bibinfo {author} {\bibfnamefont {K.}~\bibnamefont
  {Jacobs}}\ and\ \bibinfo {author} {\bibfnamefont {D.~A.}\ \bibnamefont
  {Steck}},\ }\href@noop {} {\bibfield  {journal} {\bibinfo  {journal}
  {Contemp. Phys.}\ }\textbf {\bibinfo {volume} {47}},\ \bibinfo {pages} {279}
  (\bibinfo {year} {2006})}\BibitemShut {NoStop}%
\bibitem [{\citenamefont {Di{\'o}si}\ and\ \citenamefont
  {Halliwell}(1998)}]{diosi1998coupling}%
  \BibitemOpen
  \bibfield  {author} {\bibinfo {author} {\bibfnamefont {L.}~\bibnamefont
  {Di{\'o}si}}\ and\ \bibinfo {author} {\bibfnamefont {J.~J.}\ \bibnamefont
  {Halliwell}},\ }\href@noop {} {\bibfield  {journal} {\bibinfo  {journal}
  {Phys. Rev. Lett.}\ }\textbf {\bibinfo {volume} {81}},\ \bibinfo {pages}
  {2846} (\bibinfo {year} {1998})}\BibitemShut {NoStop}%
\bibitem [{\citenamefont {Wiseman}\ and\ \citenamefont
  {Milburn}(2009)}]{wiseman2009quantum}%
  \BibitemOpen
  \bibfield  {author} {\bibinfo {author} {\bibfnamefont {H.~M.}\ \bibnamefont
  {Wiseman}}\ and\ \bibinfo {author} {\bibfnamefont {G.~J.}\ \bibnamefont
  {Milburn}},\ }\href@noop {} {\emph {\bibinfo {title} {Quantum measurement and
  control}}}\ (\bibinfo  {publisher} {Cambridge university press},\ \bibinfo
  {year} {2009})\BibitemShut {NoStop}%
\bibitem [{\citenamefont {Fox}(2006)}]{fox2006quantum}%
  \BibitemOpen
  \bibfield  {author} {\bibinfo {author} {\bibfnamefont {M.}~\bibnamefont
  {Fox}},\ }\href@noop {} {\emph {\bibinfo {title} {Quantum optics: an
  introduction}}},\ Vol.~\bibinfo {volume} {15}\ (\bibinfo  {publisher} {OUP
  Oxford},\ \bibinfo {year} {2006})\BibitemShut {NoStop}%
\bibitem [{\citenamefont {Collett}\ and\ \citenamefont
  {Gardiner}(1984)}]{collett1984squeezing}%
  \BibitemOpen
  \bibfield  {author} {\bibinfo {author} {\bibfnamefont {M.}~\bibnamefont
  {Collett}}\ and\ \bibinfo {author} {\bibfnamefont {C.}~\bibnamefont
  {Gardiner}},\ }\href@noop {} {\bibfield  {journal} {\bibinfo  {journal}
  {Phys. Rev. A}\ }\textbf {\bibinfo {volume} {30}},\ \bibinfo {pages} {1386}
  (\bibinfo {year} {1984})}\BibitemShut {NoStop}%
\bibitem [{\citenamefont {Gardiner}\ and\ \citenamefont
  {Zoller}(2004)}]{gardiner2004quantum}%
  \BibitemOpen
  \bibfield  {author} {\bibinfo {author} {\bibfnamefont {C.}~\bibnamefont
  {Gardiner}}\ and\ \bibinfo {author} {\bibfnamefont {P.}~\bibnamefont
  {Zoller}},\ }\href@noop {} {\emph {\bibinfo {title} {Quantum noise: a
  handbook of Markovian and non-Markovian quantum stochastic methods with
  applications to quantum optics}}},\ Vol.~\bibinfo {volume} {56}\ (\bibinfo
  {publisher} {Springer Science \& Business Media},\ \bibinfo {year}
  {2004})\BibitemShut {NoStop}%
\bibitem [{\citenamefont {Smithey}\ \emph {et~al.}(1993)\citenamefont
  {Smithey}, \citenamefont {Beck}, \citenamefont {Raymer},\ and\ \citenamefont
  {Faridani}}]{smithey1993measurement}%
  \BibitemOpen
  \bibfield  {author} {\bibinfo {author} {\bibfnamefont {D.}~\bibnamefont
  {Smithey}}, \bibinfo {author} {\bibfnamefont {M.}~\bibnamefont {Beck}},
  \bibinfo {author} {\bibfnamefont {M.~G.}\ \bibnamefont {Raymer}}, \ and\
  \bibinfo {author} {\bibfnamefont {A.}~\bibnamefont {Faridani}},\ }\href@noop
  {} {\bibfield  {journal} {\bibinfo  {journal} {Phys. Rev. Lett.}\ }\textbf
  {\bibinfo {volume} {70}},\ \bibinfo {pages} {1244} (\bibinfo {year}
  {1993})}\BibitemShut {NoStop}%
\bibitem [{\citenamefont {Opatrn{\`y}}\ \emph {et~al.}(1997)\citenamefont
  {Opatrn{\`y}}, \citenamefont {Welsch},\ and\ \citenamefont
  {Vogel}}]{opatrny1997homodyne}%
  \BibitemOpen
  \bibfield  {author} {\bibinfo {author} {\bibfnamefont {T.}~\bibnamefont
  {Opatrn{\`y}}}, \bibinfo {author} {\bibfnamefont {D.-G.}\ \bibnamefont
  {Welsch}}, \ and\ \bibinfo {author} {\bibfnamefont {W.}~\bibnamefont
  {Vogel}},\ }\href@noop {} {\bibfield  {journal} {\bibinfo  {journal} {Phys.
  Rev. A}\ }\textbf {\bibinfo {volume} {55}},\ \bibinfo {pages} {1416}
  (\bibinfo {year} {1997})}\BibitemShut {NoStop}%
\bibitem [{\citenamefont {Bose}\ \emph {et~al.}(1997)\citenamefont {Bose},
  \citenamefont {Jacobs},\ and\ \citenamefont {Knight}}]{bose1997preparation}%
  \BibitemOpen
  \bibfield  {author} {\bibinfo {author} {\bibfnamefont {S.}~\bibnamefont
  {Bose}}, \bibinfo {author} {\bibfnamefont {K.}~\bibnamefont {Jacobs}}, \ and\
  \bibinfo {author} {\bibfnamefont {P.}~\bibnamefont {Knight}},\ }\href@noop {}
  {\bibfield  {journal} {\bibinfo  {journal} {Phys. Rev. A}\ }\textbf {\bibinfo
  {volume} {56}},\ \bibinfo {pages} {4175} (\bibinfo {year}
  {1997})}\BibitemShut {NoStop}%
\bibitem [{\citenamefont {Nunnenkamp}\ \emph {et~al.}(2011)\citenamefont
  {Nunnenkamp}, \citenamefont {B{\o}rkje},\ and\ \citenamefont
  {Girvin}}]{nunnenkamp2011single}%
  \BibitemOpen
  \bibfield  {author} {\bibinfo {author} {\bibfnamefont {A.}~\bibnamefont
  {Nunnenkamp}}, \bibinfo {author} {\bibfnamefont {K.}~\bibnamefont
  {B{\o}rkje}}, \ and\ \bibinfo {author} {\bibfnamefont {S.~M.}\ \bibnamefont
  {Girvin}},\ }\href@noop {} {\bibfield  {journal} {\bibinfo  {journal} {Phys.
  Rev. Lett.}\ }\textbf {\bibinfo {volume} {107}},\ \bibinfo {pages} {063602}
  (\bibinfo {year} {2011})}\BibitemShut {NoStop}%
\bibitem [{\citenamefont {Hand}\ and\ \citenamefont
  {Finch}(1998)}]{hand1998analytical}%
  \BibitemOpen
  \bibfield  {author} {\bibinfo {author} {\bibfnamefont {L.~N.}\ \bibnamefont
  {Hand}}\ and\ \bibinfo {author} {\bibfnamefont {J.~D.}\ \bibnamefont
  {Finch}},\ }\href@noop {} {\emph {\bibinfo {title} {Analytical mechanics}}}\
  (\bibinfo  {publisher} {Cambridge University Press},\ \bibinfo {year}
  {1998})\BibitemShut {NoStop}%
\bibitem [{\citenamefont {Sala}\ and\ \citenamefont
  {Tufarelli}(2018)}]{sala2018exploring}%
  \BibitemOpen
  \bibfield  {author} {\bibinfo {author} {\bibfnamefont {K.}~\bibnamefont
  {Sala}}\ and\ \bibinfo {author} {\bibfnamefont {T.}~\bibnamefont
  {Tufarelli}},\ }\href@noop {} {\bibfield  {journal} {\bibinfo  {journal}
  {Sci. Rep.}\ }\textbf {\bibinfo {volume} {8}},\ \bibinfo {pages} {9157}
  (\bibinfo {year} {2018})}\BibitemShut {NoStop}%
\end{thebibliography}%

\end{document}